\documentclass[submission,copyright]{eptcs}
\providecommand{\event}{TFPIE} 
\usepackage{breakurl}             

\usepackage{savesym}
\savesymbol{amalg}

\usepackage[utf8]{inputenc}
\usepackage[T1]{fontenc}

\usepackage{mathpartir}

\newcommand{\packageGraphicx}{\usepackage{graphicx}}
\newcommand{\packageHyperref}{\usepackage{hyperref}}
\newcommand{\renewrmdefault}{\renewcommand{\rmdefault}{ptm}}
\newcommand{\packageRelsize}{\usepackage{relsize}}
\newcommand{\packageMathabx}{\usepackage{mathabx}}
\newcommand{\packageWasysym}{
  \let\leftmoon\relax \let\rightmoon\relax \let\fullmoon\relax \let\newmoon\relax \let\diameter\relax
  \usepackage{wasysym}}
\newcommand{\packageTextcomp}{\usepackage{textcomp}}
\newcommand{\packageFramed}{\usepackage{framed}}
\newcommand{\packageHyphenat}{\usepackage[htt]{hyphenat}}
\newcommand{\packageColor}{\usepackage[usenames,dvipsnames]{color}}
\newcommand{\doHypersetup}{\hypersetup{bookmarks=true,bookmarksopen=true,bookmarksnumbered=true}}
\newcommand{\packageTocstyle}{\IfFileExists{tocstyle.sty}{\usepackage{tocstyle}\usetocstyle{standard}}{}}
\newcommand{\packageCJK}{\IfFileExists{CJK.sty}{\usepackage{CJK}}{}}

\packageGraphicx
\packageHyperref
\renewrmdefault
\packageRelsize
\packageMathabx
\packageWasysym
\packageTextcomp
\packageFramed
\packageHyphenat
\packageColor
\doHypersetup
\packageTocstyle
\packageCJK

\newcommand{\BookRef}[2]{\emph{#2}}
\newcommand{\ChapRef}[2]{\SecRef{#1}{#2}}
\newcommand{\SecRef}[2]{section~#1}
\newcommand{\PartRef}[2]{part~#1}
\newcommand{\BookRefUC}[2]{\BookRef{#1}{#2}}
\newcommand{\ChapRefUC}[2]{\SecRefUC{#1}{#2}}
\newcommand{\SecRefUC}[2]{Section~#1}
\newcommand{\PartRefUC}[2]{Part~#1}

\newcommand{\BookRefLocal}[3]{\hyperref[#1]{\BookRef{#2}{#3}}}
\newcommand{\ChapRefLocal}[3]{\hyperref[#1]{\ChapRef{#2}{#3}}}
\newcommand{\SecRefLocal}[3]{\hyperref[#1]{\SecRef{#2}{#3}}}
\newcommand{\PartRefLocal}[3]{\hyperref[#1]{\PartRef{#2}{#3}}}
\newcommand{\BookRefLocalUC}[3]{\hyperref[#1]{\BookRefUC{#2}{#3}}}
\newcommand{\ChapRefLocalUC}[3]{\hyperref[#1]{\ChapRefUC{#2}{#3}}}
\newcommand{\SecRefLocalUC}[3]{\hyperref[#1]{\SecRefUC{#2}{#3}}}
\newcommand{\PartRefLocalUC}[3]{\hyperref[#1]{\PartRefUC{#2}{#3}}}

\newcommand{\BookRefUN}[1]{\BookRef{}{#1}}

\newcommand{\BookRefLocalUN}[2]{\hyperref[#1]{\BookRefUN{#2}}}

\newcommand{\Scribtexttt}[1]{{\texttt{#1}}}

\newcommand{\planetName}[1]{PLane\hspace{-0.1ex}T}

\newcommand{\Stttextmore}{{\fontencoding{T1}\selectfont>}}

\makeatletter
\def\empty@finalstrut#1{%
  \unskip\ifhmode\nobreak\fi\vrule\@width\z@\@height\z@\@depth\z@}
\def\no@strut{\global\setbox\@arstrutbox\hbox{%
    \vrule \@height\z@
           \@depth\z@
           \@width\z@}%
    \gdef\@endpbox{\empty@finalstrut\@arstrutbox\par\egroup\hfil}%
}%
\def\yes@strut{\global\setbox\@arstrutbox\hbox{%
    \vrule \@height\arraystretch \ht\strutbox
           \@depth\arraystretch \dp\strutbox
           \@width\z@}%
    \gdef\@endpbox{\@finalstrut\@arstrutbox\par\egroup\hfil}%
}%
\def\@mkpream#1{\@firstamptrue\@lastchclass6
  \let\@preamble\@empty\def\empty@preamble{\add@ins}%
  \let\protect\@unexpandable@protect
  \let\@sharp\relax\let\add@ins\relax
  \let\@startpbox\relax\let\@endpbox\relax
  \@expast{#1}%
  \expandafter\@tfor \expandafter
    \@nextchar \expandafter:\expandafter=\reserved@a\do
       {\@testpach\@nextchar
    \ifcase \@chclass \@classz \or \@classi \or \@classii \or \@classiii
      \or \@classiv \or\@classv \fi\@lastchclass\@chclass}%
  \ifcase \@lastchclass \@acol
      \or \or \@preamerr \@ne\or \@preamerr \tw@\or \or \@acol \fi}
\def\@addamp{%
  \if@firstamp
    \@firstampfalse
    \edef\empty@preamble{\add@ins}%
  \else
    \edef\@preamble{\@preamble &}%
    \edef\empty@preamble{\expandafter\noexpand\empty@preamble &\add@ins}%
  \fi}
\newif\iftw@hlines \tw@hlinesfalse
\def\@xhline{\ifx\reserved@a\hline
               \tw@hlinestrue
             \else\ifx\reserved@a\Hline
               \tw@hlinestrue
             \else
               \tw@hlinesfalse
             \fi\fi
      \iftw@hlines
        \aftergroup\do@after
      \fi
      \ifnum0=`{\fi}%
}
\def\do@after{\emptyrow[\the\doublerulesep]}
\def\emptyrow{\noalign\bgroup\@ifnextchar[\@emptyrow{\@emptyrow[\z@]}}
\def\@emptyrow[#1]{\no@strut\gdef\add@ins{\vrule \@height\z@ \@depth#1 \@width\z@}\egroup%
\empty@preamble\\
\noalign{\yes@strut\gdef\add@ins{\vrule \@height\z@ \@depth\z@ \@width\z@}}%
}
\def\tabrow#1{\noalign\bgroup\@ifnextchar[{\@tabrow{#1}}{\@tabrow{#1}[]}}
\def\@tabrow#1[#2]{\no@strut\egroup#1\ifx.#2.\\\else\\[#2]\fi\noalign{\yes@strut}}
\def\endpltstabular{\crcr\egroup\egroup \egroup}
\expandafter \let \csname endpltstabular*\endcsname = \endpltstabular
\def\pltstabular{\let\@halignto\@empty\@pltstabular}
\@namedef{pltstabular*}#1{\def\@halignto{to#1}\@pltstabular}
\def\@pltstabular{\leavevmode \bgroup \let\@acol\@tabacol
   \let\@classz\@tabclassz
   \let\@classiv\@tabclassiv \let\\\@tabularcr\@stabarray}
\def\@stabarray{\m@th\@ifnextchar[\@sarray{\@sarray[c]}}
\def\@sarray[#1]#2{%
  \bgroup
  \setbox\@arstrutbox\hbox{%
    \vrule \@height\arraystretch\ht\strutbox
           \@depth\arraystretch \dp\strutbox
           \@width\z@}%
  \@mkpream{#2}%
  \edef\@preamble{%
    \ialign \noexpand\@halignto
      \bgroup \@arstrut \@preamble \tabskip\z@skip \cr}%
  \let\@startpbox\@@startpbox \let\@endpbox\@@endpbox
  \let\tabularnewline\\%
    \let\@sharp##%
    \set@typeset@protect
    \lineskip\z@skip\baselineskip\z@skip
    \@preamble}

\makeatother

\newenvironment{bigtabular}{\begin{pltstabular}}{\end{pltstabular}}

\newlength{\stabLeft}
\newcommand{\bigtableleftpad}{\hspace{\stabLeft}}

\newenvironment{SingleColumn}{\begin{list}{}{\topsep=0pt\partopsep=0pt%
\listparindent=0pt\itemindent=0pt\labelwidth=0pt\leftmargin=0pt\rightmargin=0pt%
\itemsep=0pt\parsep=0pt}\item}{\end{list}}

\newcommand{\SCodePreSkip}{\vskip\abovedisplayskip}
\newcommand{\SCodePostSkip}{\vskip\belowdisplayskip}

\newcommand{\SVInsetPreSkip}{\vskip\abovedisplayskip}
\newcommand{\SVInsetPostSkip}{\vskip\belowdisplayskip}

\newcommand{\SAuthorSep}[1]{\qquad}
\newcommand{\SVersionBefore}[1]{Version }

\newcommand{\SNumberOfAuthors}[1]{}

\let\SOriginalthesubsection\thesubsection
\let\SOriginalthesubsubsection\thesubsubsection

\newcommand{\Ssection}[2]{\section[#1]{#2}\let\thesubsection\SOriginalthesubsection}
\newcommand{\Ssubsection}[2]{\subsection[#1]{#2}\let\thesubsubsection\SOriginalthesubsubsection}

\newcounter{GrouperTemp}

\newcommand{\Snolinkurl}[1]{\nolinkurl{#1}}

\renewcommand{\ChapRef}[2]{\SecRef{#1}{#2}}
\renewcommand{\SecRef}[2]{\S#1 ``#2''}
\renewcommand{\ChapRefUC}[2]{\SecRefUC{#1}{#2}}
\renewcommand{\SecRefUC}[2]{\SecRef{#1}{#2}}

\newcommand{\SColorize}[2]{\color{#1}{#2}}

\newcommand{\inColor}[2]{{\Scribtexttt{\SColorize{#1}{#2}}}}
\definecolor{PaleBlue}{rgb}{0.90,0.90,1.0}
\definecolor{LightGray}{rgb}{0.90,0.90,0.90}
\definecolor{CommentColor}{rgb}{0.76,0.45,0.12}
\definecolor{ParenColor}{rgb}{0.52,0.24,0.14}
\definecolor{IdentifierColor}{rgb}{0.15,0.15,0.50}
\definecolor{ResultColor}{rgb}{0.0,0.0,0.69}
\definecolor{ValueColor}{rgb}{0.13,0.55,0.13}
\definecolor{OutputColor}{rgb}{0.59,0.00,0.59}

\newcommand{\RktCmt}[1]{\inColor{CommentColor}{#1}}
\newcommand{\RktPn}[1]{\inColor{ParenColor}{#1}}

\newcommand{\RktSym}[1]{\inColor{IdentifierColor}{#1}}

\newcommand{\RktVal}[1]{\inColor{ValueColor}{#1}}

\newcommand{\RktOut}[1]{\inColor{OutputColor}{#1}}

\newcommand{\RktRdr}[1]{\inColor{black}{#1}}
\newcommand{\RktVarCol}[1]{\inColor{IdentifierColor}{#1}}
\newcommand{\RktVar}[1]{{\RktVarCol{\textsl{#1}}}}

\newenvironment{RktBlk}{}{}

\newcommand{\RBackgroundLabel}[1]{}

\newcommand{\la}{\lambda}
\newcommand{\ra}{\rightarrow}
\newcommand{\Ra}{\Rightarrow}
\newcommand{\La}{\Leftarrow}

\title{Proust: A Nano Proof Assistant}
\author{Prabhakar Ragde
\institute{Cheriton School of Computer Science\\
University of Waterloo\\
Waterloo, Ontario, Canada}
\email{plragde@uwaterloo.ca}
}
\def\titlerunning{Proust: A Nano Proof Assistant}
\def\authorrunning{P.~Ragde}

\begin{document}
\maketitle

\begin{abstract}
  Proust is a small Racket program offering rudimentary interactive
  assistance in the development of verified proofs for propositional
  and predicate logic.  It is constructed in stages, some of which are
  done by students before using it to complete proof exercises, and in
  parallel with the study of its theoretical underpinnings, including
  elements of Martin-L\"of type theory. The goal is twofold: to
  demystify some of the machinery behind full-featured proof
  assistants such as Coq and Agda, and to better integrate the study
  of formal logic with other core elements of an undergraduate
  computer science curriculum.
\end{abstract}

\section{Introduction}\label{Intro}

Most computer science programs include some exposure to logic as a
required subject. Often it is crammed into a single course on discrete
mathematics, perhaps conflated with Boolean algebra and wedged between
combinatorial counting and graph theory. At the University of
Waterloo, where I teach, we are fortunate to have CS located in a
Faculty of Mathematics, so our students take the same high-quality
math courses (discrete and continuous) taken by math
majors. Nonetheless, we have a required second-year CS course titled
Logic and Computation (henceforth L\&C).

L\&C was a relatively recent addition to our curriculum
(about fifteen years ago).
While it is a prerequisite for the data structures and
algorithms courses, that is more a matter of maturity rather than
content, and consequently the content of L\&C has tended to vary
across offerings, from ``math-style'' (Hilbert proofs, emphasis on
metamathematics) to ``SE-style'' (emphasis on program correctness).
The efficacy of these approaches is open to question, especially since
the amount of material available encourages broad, shallow treatments,
and students may tend to come away with exposure to what an individual
instructor thinks is important, rather than what might be most
beneficial for future courses or careers.

It might make more sense to deal with some of these topics as they
arise in other courses. For example, students in our first-year CS
sequence learn structural recursion on natural numbers and lists, and
this might be a place to prove properties of the code (starting with
correctness) by induction. But these properties, as well as the
subtasks involved in proofs by induction, involve manipulation of
logical statements with for-all quantifiers, and students have difficulties with
this. At least some of these difficulties stem from the fact that
for-all is a binding construct whose use has consequences
similar to the use of lambda. Proofs might better be deferred
until after students have experience with higher-order functions.

This similarity between for-all and lambda suggests another approach.
In the Curry-Howard correspondence, logical statements
(propositions or formulas)
correspond to types, and the proof of a for-all statement is a
dependently-typed function (a generalized lambda).  At first glance,
the idea of using dependent types to introduce undergraduates to logic
may seem misguided. But this approach makes use of the fact that CS
students already have experience with a formal system: a programming
language. Furthermore, our undergraduates have had a first course in
functional programming, including both theoretical and concrete use of
a substitution model (beta-reduction, though not called that or
defined in full generality) to explain the execution of programs.

Consider the interpretation of an implication. The way we prove a
statement of the form ``If T, then V'' is by assuming we have a proof
of T, and using that assumption at several points in a proof of V. If
we were then given an actual proof of T, we could substitute that for
the assumption every place it occurs, and get a self-contained proof
of V. The substitution process corresponds to the substitution of an
argument value for a parameter name everywhere in the body of a
function. Thus a proof of ``If T, then V'' corresponds to a function that
consumes a proof of T and produces a proof of V, and the use of such
an implication in a proof corresponds to function application.

To a nascent computer scientist, this is a powerful metaphor,
especially when they learn that ``If T, then V'' is usually written in
formal logic as $T\ra V$, which is also the notation for the type of a
function from $T$ to $V$.  The metaphor has more resonance than a
representation of a proof as a tree or a DAG. This is the starting
point of the several construction steps that result in Proust,
a program that both assists students in building proofs (represented
by functions) and checks their validity.
The construction process is an integral part of the L\&C course,
with some programming tasks being homework exercises.
I will describe the development of Proust, as connectives
and quantifiers are added incrementally, in sections \ref{Prop} and
\ref{Pred}, followed by discussion in section \ref{Curric} of
wider curricular issues.

I have chosen to use the Racket programming language \cite{Racket} to
implement Proust, because that is the language used by our students in
first year (plus all the reasons why we made that choice), and because
S-expressions are particularly convenient as a data representation,
but other functional languages could also be used. The design is a
synthesis from many sources, notably the OPLSS 2014 lectures of
Stephanie Weirich \cite{Weir} and her work with collaborators on the
Trellys project; blog posts by Lennart Augustsson \cite{August} and
Andrej Bauer \cite{Bauer}; papers on small dependently-typed systems
for tutorial purposes (e.g.~Altenkirch, Danielsson, L\"oh, and Oury
\cite{ADLO}; L\"oh, McBride, and Swierstra \cite{LMS}); monographs
(e.g.~Sorensen and Urzyczyn \cite{SU}); and the extensive literature
on the languages Agda \cite{Agda} and Coq \cite{Coq}.

\section{Propositional Logic}\label{Prop}

Starting with the implication metaphor discussed in section
\ref{Intro}, I derive proof rules for intuitionistic propositional
logic (normally called inference rules, but that word will be useful
elsewhere). The Proust program that checks proofs
is a straightforward implementation of these rules. Along the way,
though, I make some unconventional choices for pedagogical benefit.

From the previous discussion, we have types (logical statements, by
the Curry-Howard correspondence) constructed from variables
($A,B,C\ldots$) and implication (infix $\ra$).  We also have functions
(proof terms, by the correspondence)
with parameters (also usually called variables, $x,y,z\ldots$),
and function application.
I will use lambda-calculus notation for terms (also called expressions)
in the discussion here, but
the Proust language adapts the syntax somewhat to ease
the Racket implementation. In sequents (defined shortly)
and proof rules, I will use $T,V,W$
as type metavariables, and $f,g,a,b,c,t,v,w$ as term metavariables. $t:T$
will be the assertion that term $t$ has type $T$.

To check $\la x.t:T\ra W$, we need to check $t:W$, but we need to do
so with the knowledge that $x:T$, as $x$ is likely to occur in $t$.
This suggests maintaining a set of name-type bindings $\Gamma$
(typically called a context), yielding a three-way relation
$\Gamma\vdash t:W$ to be checked (this is a sequent). 
But checking a function application gives us some slight
trouble. To check $\Gamma\vdash f~a:W$, we need to check
$\Gamma\vdash f:T\ra W$ and $\Gamma\vdash a:T$.
Where does $T$ come from? We must infer it by looking at $f$.

\clearpage

This suggests a bidirectional approach \cite{TCoq}. I introduce the notation
$\Gamma\vdash t\La W$ to refer to the idea of type checking we started
with, namely that term $t$ has given type $W$ in context $\Gamma$. But
for type inference, I use the notation $\Gamma\vdash t\Ra T$, indicating
that in context $\Gamma$ we are able to infer that $t$ has type $T$.
To infer the type of a term that is simply a variable, we look it up
in the context.
Denoting $\Gamma \cup \{x:T\}$ as $\Gamma,x:T$, here are the
proof rules derived so far.

\begin{mathpar}
 \inferrule*[Right=(Var)]{~}{\Gamma,x:T\vdash x\Ra T}
 \and
 \inferrule*[Right=($\ra_E$)]{\Gamma,x:T\vdash t\La W}{\Gamma\vdash \la
   x.t\La T\ra W}
 \and
 \inferrule*[Right=($\ra_I$)]{\Gamma\vdash f\Ra T \ra W \\ \Gamma\vdash a \La
   T}{\Gamma\vdash f~a \Ra W}
\end{mathpar}

Rules are given names with subscripts that indicate elimination (use)
and introduction (creation), a pattern that will recur with other
logical connectives. To typecheck a term for which we don't have a
proof rule, we infer it and check that we get the same type.  We
cannot infer the type of a lambda, which means we also cannot check the
type of an immediate application of a lambda. I discuss the curricular
implications in section \ref{Curric}; for convenience, we introduce
optional type annotation on terms. Here are the additional proof rules.

\begin{mathpar}
  \inferrule*[Right=(Turn)]{\Gamma\vdash t \Ra T}{\Gamma\vdash t\La T}
  \and
  \inferrule*[Right=(Ann)]{\Gamma\vdash t\La T}{\Gamma\vdash (t:T)\Ra T}
\end{mathpar}

Using this, we can avoid both the overhead of mandatory type
annotation on term variables (as in the simply-typed lambda calculus)
and the need to discuss Hindley-Milner-style type inference.
The result is a set of syntax-directed proof rules
which have a clear implementation as a pair of mutually-recursive
procedures for checking and inferring.

Here is the initial grammar for types and terms, as the user will specify
them when using Proust.

\label{t:x28part_x28gentag_0x29x29}

\begin{bigtabular}{@{\bigtableleftpad}r@{}l@{}c@{}l@{}l@{}}
\hbox{\mbox{\hphantom{\Scribtexttt{xx}}}\RktVar{expr}} &
\hbox{\Scribtexttt{~}} &
\hbox{=} &
\hbox{\Scribtexttt{~}} &
\hbox{\RktPn{(}\RktVar{$\lambda$}\mbox{\hphantom{\Scribtexttt{x}}}\RktVar{x}\mbox{\hphantom{\Scribtexttt{x}}}\RktVar{={\Stttextmore}}\mbox{\hphantom{\Scribtexttt{x}}}\RktVar{expr}\RktPn{)}} \\
\hbox{\Scribtexttt{~}} &
\hbox{\Scribtexttt{~}} &
\hbox{$|$} &
\hbox{\Scribtexttt{~}} &
\hbox{\RktPn{(}\RktVar{expr}\mbox{\hphantom{\Scribtexttt{x}}}\RktVar{expr}\RktPn{)}} \\
\hbox{\Scribtexttt{~}} &
\hbox{\Scribtexttt{~}} &
\hbox{$|$} &
\hbox{\Scribtexttt{~}} &
\hbox{\RktPn{(}\RktVar{expr}\mbox{\hphantom{\Scribtexttt{x}}}\RktVar{{\hbox{\texttt{:}}}}\mbox{\hphantom{\Scribtexttt{x}}}\RktVar{type}\RktPn{)}} \\
\hbox{\Scribtexttt{~}} &
\hbox{\Scribtexttt{~}} &
\hbox{$|$} &
\hbox{\Scribtexttt{~}} &
\hbox{\RktVar{x}} \\
\hbox{\Scribtexttt{~}} &
\hbox{\Scribtexttt{~}} &
\hbox{\Scribtexttt{~}} &
\hbox{\Scribtexttt{~}} &
\hbox{\Scribtexttt{~}} \\
\hbox{\mbox{\hphantom{\Scribtexttt{xx}}}\RktVar{type}} &
\hbox{\Scribtexttt{~}} &
\hbox{=} &
\hbox{\Scribtexttt{~}} &
\hbox{\RktPn{(}\RktVar{type}\mbox{\hphantom{\Scribtexttt{x}}}\RktVar{{-}{\Stttextmore}}\mbox{\hphantom{\Scribtexttt{x}}}\RktVar{type}\RktPn{)}} \\
\hbox{\Scribtexttt{~}} &
\hbox{\Scribtexttt{~}} &
\hbox{$|$} &
\hbox{\Scribtexttt{~}} &
\hbox{\RktVar{X}}\end{bigtabular}

The program does not enforce the lexical restrictions on
choice of variables described above, or even the convention that
parameters are lower-case while type variables are upper-case;
any symbols may be used.
The program parses user input into an AST built with Racket structures
(records), using Racket's built-in pattern matching.

~

\label{t:x28part_x28gentag_0x29x29}

\begin{RktBlk}\begin{SingleColumn}\RktPn{(}\RktSym{struct}\mbox{\hphantom{\Scribtexttt{x}}}\RktSym{Lam}\mbox{\hphantom{\Scribtexttt{x}}}\RktPn{(}\RktSym{var}\mbox{\hphantom{\Scribtexttt{x}}}\RktSym{body}\RktPn{)}\RktPn{)}

\RktPn{(}\RktSym{struct}\mbox{\hphantom{\Scribtexttt{x}}}\RktSym{App}\mbox{\hphantom{\Scribtexttt{x}}}\RktPn{(}\RktSym{rator}\mbox{\hphantom{\Scribtexttt{x}}}\RktSym{rand}\RktPn{)}\RktPn{)}

\mbox{\hphantom{\Scribtexttt{x}}}

\RktPn{(}\RktSym{struct}\mbox{\hphantom{\Scribtexttt{x}}}\RktSym{Arrow}\mbox{\hphantom{\Scribtexttt{x}}}\RktPn{(}\RktSym{domain}\mbox{\hphantom{\Scribtexttt{x}}}\RktSym{range}\RktPn{)}\RktPn{)}\end{SingleColumn}\end{RktBlk}

\begin{minipage}[t]{0.5\textwidth}
\label{t:x28part_x28gentag_0x29x29}

\begin{RktBlk}\begin{SingleColumn}\RktCmt{;}\RktCmt{~}\RktCmt{parse{-}expr {\hbox{\texttt{:}}} sexp {-}{\Stttextmore} Expr}

\mbox{\hphantom{\Scribtexttt{x}}}

\RktPn{(}\RktSym{define}\mbox{\hphantom{\Scribtexttt{x}}}\RktPn{(}\RktSym{parse{-}expr}\mbox{\hphantom{\Scribtexttt{x}}}\RktSym{s}\RktPn{)}

\mbox{\hphantom{\Scribtexttt{xx}}}\RktPn{(}\RktSym{match}\mbox{\hphantom{\Scribtexttt{x}}}\RktSym{s}

\mbox{\hphantom{\Scribtexttt{xxxx}}}\RktPn{[}\RktVal{{\textasciigrave}}\RktVal{(}\RktVal{$\lambda$}\mbox{\hphantom{\Scribtexttt{x}}}\RktRdr{,}\RktPn{(}\RktSym{{\hbox{\texttt{?}}}}\mbox{\hphantom{\Scribtexttt{x}}}\RktSym{symbol{\hbox{\texttt{?}}}}\mbox{\hphantom{\Scribtexttt{x}}}\RktSym{x}\RktPn{)}\mbox{\hphantom{\Scribtexttt{x}}}\RktVal{={\Stttextmore}}\mbox{\hphantom{\Scribtexttt{x}}}\RktRdr{,}\RktSym{e}\RktVal{)}

\mbox{\hphantom{\Scribtexttt{xxxxxxx}}}\RktPn{(}\RktSym{Lam}\mbox{\hphantom{\Scribtexttt{x}}}\RktSym{x}\mbox{\hphantom{\Scribtexttt{x}}}\RktPn{(}\RktSym{parse{-}expr}\mbox{\hphantom{\Scribtexttt{x}}}\RktSym{e}\RktPn{)}\RktPn{)}\RktPn{]}

\mbox{\hphantom{\Scribtexttt{xxxx}}}\RktPn{[}\RktVal{{\textasciigrave}}\RktVal{(}\RktRdr{,}\RktSym{e0}\mbox{\hphantom{\Scribtexttt{x}}}\RktRdr{,}\RktSym{e1}\RktVal{)}

\mbox{\hphantom{\Scribtexttt{xxxxxxx}}}\RktPn{(}\RktSym{App}\mbox{\hphantom{\Scribtexttt{x}}}\RktPn{(}\RktSym{parse{-}expr}\mbox{\hphantom{\Scribtexttt{x}}}\RktSym{e0}\RktPn{)}

\mbox{\hphantom{\Scribtexttt{xxxxxxxxxxxx}}}\RktPn{(}\RktSym{parse{-}expr}\mbox{\hphantom{\Scribtexttt{x}}}\RktSym{e1}\RktPn{)}\RktPn{)}\RktPn{]}

\mbox{\hphantom{\Scribtexttt{xxxx}}}\RktPn{[}\RktPn{(}\RktSym{{\hbox{\texttt{?}}}}\mbox{\hphantom{\Scribtexttt{x}}}\RktSym{symbol{\hbox{\texttt{?}}}}\mbox{\hphantom{\Scribtexttt{x}}}\RktSym{x}\RktPn{)}\mbox{\hphantom{\Scribtexttt{x}}}\RktSym{x}\RktPn{]}\RktPn{)}\RktPn{)}\end{SingleColumn}\end{RktBlk}
\end{minipage}
\begin{minipage}[t]{0.5\textwidth}
\label{t:x28part_x28gentag_0x29x29}

\begin{RktBlk}\begin{SingleColumn}\RktCmt{;}\RktCmt{~}\RktCmt{parse{-}type {\hbox{\texttt{:}}} sexp {-}{\Stttextmore} Type}

\mbox{\hphantom{\Scribtexttt{x}}}

\RktPn{(}\RktSym{define}\mbox{\hphantom{\Scribtexttt{x}}}\RktPn{(}\RktSym{parse{-}type}\mbox{\hphantom{\Scribtexttt{x}}}\RktSym{t}\RktPn{)}

\mbox{\hphantom{\Scribtexttt{xx}}}\RktPn{(}\RktSym{match}\mbox{\hphantom{\Scribtexttt{x}}}\RktSym{t}

\mbox{\hphantom{\Scribtexttt{xxxx}}}\RktPn{[}\RktVal{{\textasciigrave}}\RktVal{(}\RktRdr{,}\RktSym{t1}\mbox{\hphantom{\Scribtexttt{x}}}\RktVal{\mbox{{-}{\Stttextmore}}}\mbox{\hphantom{\Scribtexttt{x}}}\RktRdr{,}\RktSym{t2}\RktVal{)}

\mbox{\hphantom{\Scribtexttt{xxxxxxxx}}}\RktPn{(}\RktSym{Arrow}\mbox{\hphantom{\Scribtexttt{x}}}\RktPn{(}\RktSym{parse{-}type}\mbox{\hphantom{\Scribtexttt{x}}}\RktSym{t1}\RktPn{)}

\mbox{\hphantom{\Scribtexttt{xxxxxxxxxxxxxxx}}}\RktPn{(}\RktSym{parse{-}type}\mbox{\hphantom{\Scribtexttt{x}}}\RktSym{t2}\RktPn{)}\RktPn{)}\RktPn{]}

\mbox{\hphantom{\Scribtexttt{xxxx}}}\RktPn{[}\RktPn{(}\RktSym{{\hbox{\texttt{?}}}}\mbox{\hphantom{\Scribtexttt{x}}}\RktSym{symbol{\hbox{\texttt{?}}}}\mbox{\hphantom{\Scribtexttt{x}}}\RktSym{X}\RktPn{)}\mbox{\hphantom{\Scribtexttt{x}}}\RktSym{X}\RktPn{]}

\mbox{\hphantom{\Scribtexttt{xxxx}}}\RktPn{[}\RktSym{else}\mbox{\hphantom{\Scribtexttt{x}}}\RktPn{(}\RktSym{error}\mbox{\hphantom{\Scribtexttt{x}}}\RktSym{{\hbox{\texttt{.}}}{\hbox{\texttt{.}}}{\hbox{\texttt{.}}}}\RktPn{)}\RktPn{]}\RktPn{)}\RktPn{)}\end{SingleColumn}\end{RktBlk}
\end{minipage}

~

I have elided the error message in \RktSym{parse{-}type}.  When the
type and term languages are extended with new constructs in order to
deal with additional logical connectives, students can implement the
requisite parser extensions themselves.  The code for type checking
and inferring follows a similar pattern, guided by the grammar and the
proof rules. Contexts are represented as association lists.

~

\label{t:x28part_x28gentag_0x29x29}

\begin{RktBlk}\begin{SingleColumn}\RktCmt{;}\RktCmt{~}\RktCmt{type{-}check {\hbox{\texttt{:}}} Context Expr Type {-}{\Stttextmore} boolean}

\RktCmt{;}\RktCmt{~}\RktCmt{produces true if expr has type t in context ctx (or error if not)}

\mbox{\hphantom{\Scribtexttt{x}}}

\RktPn{(}\RktSym{define}\mbox{\hphantom{\Scribtexttt{x}}}\RktPn{(}\RktSym{type{-}check}\mbox{\hphantom{\Scribtexttt{x}}}\RktSym{ctx}\mbox{\hphantom{\Scribtexttt{x}}}\RktSym{expr}\mbox{\hphantom{\Scribtexttt{x}}}\RktSym{type}\RktPn{)}

\mbox{\hphantom{\Scribtexttt{xx}}}\RktPn{(}\RktSym{match}\mbox{\hphantom{\Scribtexttt{x}}}\RktSym{expr}

\mbox{\hphantom{\Scribtexttt{xxxx}}}\RktPn{[}\RktPn{(}\RktSym{Lam}\mbox{\hphantom{\Scribtexttt{x}}}\RktSym{x}\mbox{\hphantom{\Scribtexttt{x}}}\RktSym{t}\RktPn{)}

\mbox{\hphantom{\Scribtexttt{xxxxxxx}}}\RktPn{(}\RktSym{match}\mbox{\hphantom{\Scribtexttt{x}}}\RktSym{type}

\mbox{\hphantom{\Scribtexttt{xxxxxxxxx}}}\RktPn{[}\RktPn{(}\RktSym{Arrow}\mbox{\hphantom{\Scribtexttt{x}}}\RktSym{tt}\mbox{\hphantom{\Scribtexttt{x}}}\RktSym{tw}\RktPn{)}\mbox{\hphantom{\Scribtexttt{x}}}\RktPn{(}\RktSym{type{-}check}\mbox{\hphantom{\Scribtexttt{x}}}\RktPn{(}\RktSym{cons}\mbox{\hphantom{\Scribtexttt{x}}}\RktVal{{\textasciigrave}}\RktVal{(}\RktRdr{,}\RktSym{x}\mbox{\hphantom{\Scribtexttt{x}}}\RktRdr{,}\RktSym{tt}\RktVal{)}\mbox{\hphantom{\Scribtexttt{x}}}\RktSym{ctx}\RktPn{)}\mbox{\hphantom{\Scribtexttt{x}}}\RktSym{t}\mbox{\hphantom{\Scribtexttt{x}}}\RktSym{tw}\RktPn{)}\RktPn{]}

\mbox{\hphantom{\Scribtexttt{xxxxxxxxx}}}\RktPn{[}\RktSym{else}\mbox{\hphantom{\Scribtexttt{x}}}\RktPn{(}\RktSym{cannot{-}check}\mbox{\hphantom{\Scribtexttt{x}}}\RktSym{ctx}\mbox{\hphantom{\Scribtexttt{x}}}\RktSym{expr}\mbox{\hphantom{\Scribtexttt{x}}}\RktSym{type}\RktPn{)}\RktPn{]}\RktPn{)}\RktPn{]}

\mbox{\hphantom{\Scribtexttt{xxxx}}}\RktPn{[}\RktSym{else}\mbox{\hphantom{\Scribtexttt{x}}}\RktPn{(}\RktSym{if}\mbox{\hphantom{\Scribtexttt{x}}}\RktPn{(}\RktSym{equal{\hbox{\texttt{?}}}}\mbox{\hphantom{\Scribtexttt{x}}}\RktPn{(}\RktSym{type{-}infer}\mbox{\hphantom{\Scribtexttt{x}}}\RktSym{ctx}\mbox{\hphantom{\Scribtexttt{x}}}\RktSym{expr}\RktPn{)}\mbox{\hphantom{\Scribtexttt{x}}}\RktSym{type}\RktPn{)}\mbox{\hphantom{\Scribtexttt{x}}}\RktSym{true}\mbox{\hphantom{\Scribtexttt{x}}}\RktPn{(}\RktSym{cannot{-}check}\mbox{\hphantom{\Scribtexttt{x}}}\RktSym{ctx}\mbox{\hphantom{\Scribtexttt{x}}}\RktSym{expr}\mbox{\hphantom{\Scribtexttt{x}}}\RktSym{type}\RktPn{)}\RktPn{)}\RktPn{]}\RktPn{)}\RktPn{)}

\mbox{\hphantom{\Scribtexttt{x}}}

\RktCmt{;}\RktCmt{~}\RktCmt{type{-}infer {\hbox{\texttt{:}}} Context Expr {-}{\Stttextmore} Type}

\RktCmt{;}\RktCmt{~}\RktCmt{produces type of expr in context ctx (or error if can}\RktCmt{{\textquotesingle}}\RktCmt{t)}

\mbox{\hphantom{\Scribtexttt{x}}}

\RktPn{(}\RktSym{define}\mbox{\hphantom{\Scribtexttt{x}}}\RktPn{(}\RktSym{type{-}infer}\mbox{\hphantom{\Scribtexttt{x}}}\RktSym{ctx}\mbox{\hphantom{\Scribtexttt{x}}}\RktSym{expr}\RktPn{)}

\mbox{\hphantom{\Scribtexttt{xx}}}\RktPn{(}\RktSym{match}\mbox{\hphantom{\Scribtexttt{x}}}\RktSym{expr}

\mbox{\hphantom{\Scribtexttt{xxxx}}}\RktPn{[}\RktPn{(}\RktSym{Lam}\mbox{\hphantom{\Scribtexttt{x}}}\RktSym{{\char`\_}}\mbox{\hphantom{\Scribtexttt{x}}}\RktSym{{\char`\_}}\RktPn{)}\mbox{\hphantom{\Scribtexttt{x}}}\RktPn{(}\RktSym{cannot{-}infer}\mbox{\hphantom{\Scribtexttt{x}}}\RktSym{ctx}\mbox{\hphantom{\Scribtexttt{x}}}\RktSym{expr}\RktPn{)}\RktPn{]}

\mbox{\hphantom{\Scribtexttt{xxxx}}}\RktPn{[}\RktPn{(}\RktSym{Ann}\mbox{\hphantom{\Scribtexttt{x}}}\RktSym{e}\mbox{\hphantom{\Scribtexttt{x}}}\RktSym{t}\RktPn{)}\mbox{\hphantom{\Scribtexttt{x}}}\RktPn{(}\RktSym{type{-}check}\mbox{\hphantom{\Scribtexttt{x}}}\RktSym{ctx}\mbox{\hphantom{\Scribtexttt{x}}}\RktSym{e}\mbox{\hphantom{\Scribtexttt{x}}}\RktSym{t}\RktPn{)}\mbox{\hphantom{\Scribtexttt{x}}}\RktSym{t}\RktPn{]}

\mbox{\hphantom{\Scribtexttt{xxxx}}}\RktPn{[}\RktPn{(}\RktSym{App}\mbox{\hphantom{\Scribtexttt{x}}}\RktSym{f}\mbox{\hphantom{\Scribtexttt{x}}}\RktSym{a}\RktPn{)}

\mbox{\hphantom{\Scribtexttt{xxxxxxx}}}\RktPn{(}\RktSym{define}\mbox{\hphantom{\Scribtexttt{x}}}\RktSym{tf}\mbox{\hphantom{\Scribtexttt{x}}}\RktPn{(}\RktSym{type{-}infer}\mbox{\hphantom{\Scribtexttt{x}}}\RktSym{ctx}\mbox{\hphantom{\Scribtexttt{x}}}\RktSym{f}\RktPn{)}\RktPn{)}

\mbox{\hphantom{\Scribtexttt{xxxxxxxx}}}\RktPn{(}\RktSym{match}\mbox{\hphantom{\Scribtexttt{x}}}\RktSym{tf}

\mbox{\hphantom{\Scribtexttt{xxxxxxxxx}}}\RktPn{[}\RktPn{(}\RktSym{Arrow}\mbox{\hphantom{\Scribtexttt{x}}}\RktSym{tt}\mbox{\hphantom{\Scribtexttt{x}}}\RktSym{tw}\RktPn{)}\mbox{\hphantom{\Scribtexttt{x}}}\RktPn{\#{\hbox{\texttt{:}}}when}\mbox{\hphantom{\Scribtexttt{x}}}\RktPn{(}\RktSym{type{-}check}\mbox{\hphantom{\Scribtexttt{x}}}\RktSym{ctx}\mbox{\hphantom{\Scribtexttt{x}}}\RktSym{a}\mbox{\hphantom{\Scribtexttt{x}}}\RktSym{tt}\RktPn{)}\mbox{\hphantom{\Scribtexttt{x}}}\RktSym{tw}\RktPn{]}

\mbox{\hphantom{\Scribtexttt{xxxxxxxxx}}}\RktPn{[}\RktSym{else}\mbox{\hphantom{\Scribtexttt{x}}}\RktPn{(}\RktSym{cannot{-}infer}\mbox{\hphantom{\Scribtexttt{x}}}\RktSym{ctx}\mbox{\hphantom{\Scribtexttt{x}}}\RktSym{expr}\RktPn{)}\RktPn{]}\RktPn{)}\RktPn{]}

\mbox{\hphantom{\Scribtexttt{xxxx}}}\RktPn{[}\RktPn{(}\RktSym{{\hbox{\texttt{?}}}}\mbox{\hphantom{\Scribtexttt{x}}}\RktSym{symbol{\hbox{\texttt{?}}}}\mbox{\hphantom{\Scribtexttt{x}}}\RktSym{x}\RktPn{)}

\mbox{\hphantom{\Scribtexttt{xxxxxxx}}}\RktPn{(}\RktSym{cond}

\mbox{\hphantom{\Scribtexttt{xxxxxxxxx}}}\RktPn{[}\RktPn{(}\RktSym{assoc}\mbox{\hphantom{\Scribtexttt{x}}}\RktSym{x}\mbox{\hphantom{\Scribtexttt{x}}}\RktSym{ctx}\RktPn{)}\mbox{\hphantom{\Scribtexttt{x}}}\RktSym{={\Stttextmore}}\mbox{\hphantom{\Scribtexttt{x}}}\RktSym{second}\RktPn{]}

\mbox{\hphantom{\Scribtexttt{xxxxxxxxx}}}\RktPn{[}\RktSym{else}\mbox{\hphantom{\Scribtexttt{x}}}\RktPn{(}\RktSym{cannot{-}infer}\mbox{\hphantom{\Scribtexttt{x}}}\RktSym{ctx}\mbox{\hphantom{\Scribtexttt{x}}}\RktSym{expr}\RktPn{)}\RktPn{]}\RktPn{)}\RktPn{]}\RktPn{)}\RktPn{)}

\mbox{\hphantom{\Scribtexttt{x}}}

\RktCmt{;}\RktCmt{~}\RktCmt{a simple way to test{\hbox{\texttt{:}}} a proof term annotated with expected type}

\RktPn{(}\RktSym{define}\mbox{\hphantom{\Scribtexttt{x}}}\RktPn{(}\RktSym{check{-}proof}\mbox{\hphantom{\Scribtexttt{x}}}\RktSym{p}\RktPn{)}\mbox{\hphantom{\Scribtexttt{x}}}\RktPn{(}\RktSym{type{-}infer}\mbox{\hphantom{\Scribtexttt{x}}}\RktSym{empty}\mbox{\hphantom{\Scribtexttt{x}}}\RktPn{(}\RktSym{parse{-}expr}\mbox{\hphantom{\Scribtexttt{x}}}\RktSym{p}\RktPn{)}\RktPn{)}\mbox{\hphantom{\Scribtexttt{x}}}\RktSym{true}\RktPn{)}\end{SingleColumn}\end{RktBlk}

The \RktSym{cannot{-}check/infer} procedures
raise an error whose message includes
pretty-printed representations of useful information (context, term,
type); the pretty-printers use matching, and again can be
written/extended by students.

We have a type checker for the implicational fragment of
intuitionistic propositional logic. The size of the code base thus far:
4 lines for datatypes, 12 lines for parsing, 21 lines for
checking/inference, one line for a test function, and not shown here,
16 lines for pretty-printing and 6 lines for error handling,
for a total of 60 lines.

This program lets us check simple proofs such as

~

\noindent\RktSym{(check{-}proof '(($\lambda$ x => ($\lambda$ y => (y x))) : (A
  {-}> ((A {-}> B) {-}> B))))}

~

\noindent but there are also more interesting things we can check,
such as the proof of the transitivity of implication, which is the
function composition operator:

\[\lambda f .\lambda g .\lambda x . f~ (g~ x) :
(B \ra C) \ra ((A \ra B) \ra (A \ra C))\]

Because this is intuitionistic logic, there are some things we might
expect to be able to prove at this point, but cannot, such as Peirce's
Law: $((A \ra B) \ra A) \ra A$. In section \ref{Curric}, I discuss how
one might deal with this in the classroom.

I have not yet delivered on the promise of an interactive proof
assistant. To this end we make use of DrRacket's REPL
(read-evaluate-print loop), as well as its support for hash tables.
We introduce the notion of holes (or goals), which are unfinished
parts of the proof. A hole is a term represented by \RktSym{?} in the
term language and by a \RktSym{Hole} structure in the AST with a field
for a number. A hole typechecks with any provided type. In Proust, we define
state variables for the current expression, goal table (a hash table
mapping goal number to required type), and a counter
to number new holes, handled as simply as possible.

~

\label{t:x28part_x28gentag_0x29x29}

\begin{RktBlk}\begin{SingleColumn}\RktPn{(}\RktSym{define}\mbox{\hphantom{\Scribtexttt{x}}}\RktSym{current{-}expr}\mbox{\hphantom{\Scribtexttt{x}}}\RktVal{\#f}\RktPn{)}

\RktPn{(}\RktSym{define}\mbox{\hphantom{\Scribtexttt{x}}}\RktSym{goal{-}table}\mbox{\hphantom{\Scribtexttt{x}}}\RktPn{(}\RktSym{make{-}hash}\RktPn{)}\RktPn{)}

\RktPn{(}\RktSym{define}\mbox{\hphantom{\Scribtexttt{x}}}\RktSym{hole{-}ctr}\mbox{\hphantom{\Scribtexttt{x}}}\RktVal{0}\RktPn{)}

\RktPn{(}\RktSym{define}\mbox{\hphantom{\Scribtexttt{x}}}\RktPn{(}\RktSym{use{-}hole{-}ctr}\RktPn{)}\mbox{\hphantom{\Scribtexttt{x}}}\RktPn{(}\RktSym{begin0}\mbox{\hphantom{\Scribtexttt{x}}}\RktSym{hole{-}ctr}\mbox{\hphantom{\Scribtexttt{x}}}\RktPn{(}\RktSym{set{\hbox{\texttt{!}}}}\mbox{\hphantom{\Scribtexttt{x}}}\RktSym{hole{-}ctr}\mbox{\hphantom{\Scribtexttt{x}}}\RktPn{(}\RktSym{add1}\mbox{\hphantom{\Scribtexttt{x}}}\RktSym{hole{-}ctr}\RktPn{)}\RktPn{)}\RktPn{)}\RktPn{)}

\RktPn{(}\RktSym{define}\mbox{\hphantom{\Scribtexttt{x}}}\RktPn{(}\RktSym{print{-}task}\RktPn{)}\mbox{\hphantom{\Scribtexttt{x}}}\RktPn{(}\RktSym{printf}\mbox{\hphantom{\Scribtexttt{x}}}\RktVal{"$\sim$a{\char`\\}n"}\mbox{\hphantom{\Scribtexttt{x}}}\RktPn{(}\RktSym{pretty{-}print{-}expr}\mbox{\hphantom{\Scribtexttt{x}}}\RktSym{current{-}expr}\RktPn{)}\RktPn{)}\RktPn{)}\end{SingleColumn}\end{RktBlk}

~

The \RktSym{set{-}task!} procedure initializes state variables for a
new task (described as an S-expression). A typical application is
\RktPn{(}\RktSym{set{-}task!} \RktVal{{\textquotesingle}(? : T)}\RktPn{)},
where \RktSym{T} is a logical statement to be proved.

~

\label{t:x28part_x28gentag_0x29x29}

\begin{RktBlk}\begin{SingleColumn}\RktPn{(}\RktSym{define}\mbox{\hphantom{\Scribtexttt{x}}}\RktPn{(}\RktSym{set{-}task{\hbox{\texttt{!}}}}\mbox{\hphantom{\Scribtexttt{x}}}\RktSym{s}\RktPn{)}

\mbox{\hphantom{\Scribtexttt{xx}}}\RktPn{(}\RktSym{set{\hbox{\texttt{!}}}}\mbox{\hphantom{\Scribtexttt{x}}}\RktSym{goal{-}table}\mbox{\hphantom{\Scribtexttt{x}}}\RktPn{(}\RktSym{make{-}hash}\RktPn{)}\RktPn{)}

\mbox{\hphantom{\Scribtexttt{xx}}}\RktPn{(}\RktSym{set{\hbox{\texttt{!}}}}\mbox{\hphantom{\Scribtexttt{x}}}\RktSym{hole{-}ctr}\mbox{\hphantom{\Scribtexttt{x}}}\RktVal{0}\RktPn{)}

\mbox{\hphantom{\Scribtexttt{xx}}}\RktPn{(}\RktSym{define}\mbox{\hphantom{\Scribtexttt{x}}}\RktSym{e}\mbox{\hphantom{\Scribtexttt{x}}}\RktPn{(}\RktSym{parse{-}expr}\mbox{\hphantom{\Scribtexttt{x}}}\RktSym{s}\RktPn{)}\RktPn{)}

\mbox{\hphantom{\Scribtexttt{xx}}}\RktPn{(}\RktSym{match}\mbox{\hphantom{\Scribtexttt{x}}}\RktSym{e}

\mbox{\hphantom{\Scribtexttt{xxxx}}}\RktPn{[}\RktPn{(}\RktSym{Ann}\mbox{\hphantom{\Scribtexttt{x}}}\RktSym{{\char`\_}}\mbox{\hphantom{\Scribtexttt{x}}}\RktSym{{\char`\_}}\RktPn{)}\mbox{\hphantom{\Scribtexttt{x}}}\RktPn{(}\RktSym{set{\hbox{\texttt{!}}}}\mbox{\hphantom{\Scribtexttt{x}}}\RktSym{current{-}expr}\mbox{\hphantom{\Scribtexttt{x}}}\RktSym{e}\RktPn{)}\mbox{\hphantom{\Scribtexttt{x}}}\RktPn{(}\RktSym{type{-}infer}\mbox{\hphantom{\Scribtexttt{x}}}\RktSym{empty}\mbox{\hphantom{\Scribtexttt{x}}}\RktSym{e}\RktPn{)}\RktPn{]}

\mbox{\hphantom{\Scribtexttt{xxxx}}}\RktPn{[}\RktSym{else}\mbox{\hphantom{\Scribtexttt{x}}}\RktPn{(}\RktSym{error}\mbox{\hphantom{\Scribtexttt{x}}}\RktVal{"task must be of form (term {\hbox{\texttt{:}}} type)"}\RktPn{)}\RktPn{]}\RktPn{)}

\mbox{\hphantom{\Scribtexttt{xx}}}\RktPn{(}\RktSym{printf}\mbox{\hphantom{\Scribtexttt{x}}}\RktVal{"Task is now{\char`\\}n"}\RktPn{)}

\mbox{\hphantom{\Scribtexttt{xx}}}\RktPn{(}\RktSym{print{-}task}\RktPn{)}\RktPn{)}\end{SingleColumn}\end{RktBlk}

\clearpage

The \RktSym{refine} procedure refines goal $n$ with expression $s$,
but makes sure it typechecks first. If it does, it removes goal $n$
and rewrites the current expression to replace that goal with $s$,
using \RktSym{replace-goal-with}, which does straightforward structural
recursion on the AST (the goal being at a single leaf).

~

\label{t:x28part_x28gentag_0x29x29}

\begin{RktBlk}\begin{SingleColumn}\RktPn{(}\RktSym{define}\mbox{\hphantom{\Scribtexttt{x}}}\RktPn{(}\RktSym{refine}\mbox{\hphantom{\Scribtexttt{x}}}\RktSym{n}\mbox{\hphantom{\Scribtexttt{x}}}\RktSym{s}\RktPn{)}

\mbox{\hphantom{\Scribtexttt{xx}}}\RktPn{(}\RktSym{match{-}define}\mbox{\hphantom{\Scribtexttt{x}}}\RktPn{(}\RktSym{list}\mbox{\hphantom{\Scribtexttt{x}}}\RktSym{t}\mbox{\hphantom{\Scribtexttt{x}}}\RktSym{ctx}\RktPn{)}

\mbox{\hphantom{\Scribtexttt{xxxx}}}\RktPn{(}\RktSym{hash{-}ref}\mbox{\hphantom{\Scribtexttt{x}}}\RktSym{goal{-}table}\mbox{\hphantom{\Scribtexttt{x}}}\RktSym{n}\mbox{\hphantom{\Scribtexttt{x}}}\RktPn{(}\RktSym{lambda}\mbox{\hphantom{\Scribtexttt{x}}}\RktPn{(}\RktPn{)}\mbox{\hphantom{\Scribtexttt{x}}}\RktPn{(}\RktSym{error}\mbox{\hphantom{\Scribtexttt{x}}}\RktVal{{\textquotesingle}}\RktVal{refine}\mbox{\hphantom{\Scribtexttt{x}}}\RktVal{"no goal numbered $\sim$a"}\mbox{\hphantom{\Scribtexttt{x}}}\RktSym{n}\RktPn{)}\RktPn{)}\RktPn{)}\RktPn{)}

\mbox{\hphantom{\Scribtexttt{xx}}}\RktPn{(}\RktSym{define}\mbox{\hphantom{\Scribtexttt{x}}}\RktSym{e}\mbox{\hphantom{\Scribtexttt{x}}}\RktPn{(}\RktSym{parse{-}expr}\mbox{\hphantom{\Scribtexttt{x}}}\RktSym{s}\RktPn{)}\RktPn{)}

\mbox{\hphantom{\Scribtexttt{xx}}}\RktPn{(}\RktSym{type{-}check}\mbox{\hphantom{\Scribtexttt{x}}}\RktSym{ctx}\mbox{\hphantom{\Scribtexttt{x}}}\RktSym{e}\mbox{\hphantom{\Scribtexttt{x}}}\RktSym{t}\RktPn{)}

\mbox{\hphantom{\Scribtexttt{xx}}}\RktPn{(}\RktSym{hash{-}remove{\hbox{\texttt{!}}}}\mbox{\hphantom{\Scribtexttt{x}}}\RktSym{goal{-}table}\mbox{\hphantom{\Scribtexttt{x}}}\RktSym{n}\RktPn{)}

\mbox{\hphantom{\Scribtexttt{xx}}}\RktPn{(}\RktSym{set{\hbox{\texttt{!}}}}\mbox{\hphantom{\Scribtexttt{x}}}\RktSym{current{-}expr}\mbox{\hphantom{\Scribtexttt{x}}}\RktPn{(}\RktSym{replace{-}goal{-}with}\mbox{\hphantom{\Scribtexttt{x}}}\RktSym{n}\mbox{\hphantom{\Scribtexttt{x}}}\RktSym{e}\mbox{\hphantom{\Scribtexttt{x}}}\RktSym{current{-}expr}\RktPn{)}\RktPn{)}

\mbox{\hphantom{\Scribtexttt{xx}}}\RktPn{(}\RktSym{printf}\mbox{\hphantom{\Scribtexttt{x}}}\RktVal{"Task is now{\char`\\}n"}\mbox{\hphantom{\Scribtexttt{x}}}\RktPn{(}\RktSym{format}\mbox{\hphantom{\Scribtexttt{x}}}\RktVal{"$\sim$a goal$\sim$a"}\mbox{\hphantom{\Scribtexttt{x}}}\RktSym{ngoals}\mbox{\hphantom{\Scribtexttt{x}}}\RktPn{(}\RktSym{if}\mbox{\hphantom{\Scribtexttt{x}}}\RktPn{(}\RktSym{=}\mbox{\hphantom{\Scribtexttt{x}}}\RktSym{ngoals}\mbox{\hphantom{\Scribtexttt{x}}}\RktVal{1}\RktPn{)}\mbox{\hphantom{\Scribtexttt{x}}}\RktVal{""}\mbox{\hphantom{\Scribtexttt{x}}}\RktVal{"s"}\RktPn{)}\RktPn{)}\RktPn{)}

\mbox{\hphantom{\Scribtexttt{xx}}}\RktPn{(}\RktSym{print{-}task}\RktPn{)}\RktPn{)}\end{SingleColumn}\end{RktBlk}

~

With still under a hundred lines of code, we now have the
capability of interactions like the following, which proves the
``transitivity of implication'' result described above. The
$>$ character is the REPL prompt. The interaction is crude, but
genuinely helpful (more so with simple commands to pretty-print
information about the goal and its context),
and foreshadows similar ideas in interactions with
Agda and Coq.

~

{\parindent=0pt
\label{t:x28part_x28gentag_0x29x29}

$>$ \RktPn{(}\RktSym{set{-}task{\hbox{\texttt{!}}}}\Scribtexttt{ }\RktVal{{\textquotesingle}}\RktVal{(}\RktVal{{\hbox{\texttt{?}}}}\Scribtexttt{ }\RktVal{{\hbox{\texttt{:}}}}\Scribtexttt{ }\RktVal{(}\RktVal{(}\RktVal{B}\Scribtexttt{ }\RktVal{\mbox{{-}{\Stttextmore}}}\Scribtexttt{ }\RktVal{C}\RktVal{)}\Scribtexttt{ }\RktVal{\mbox{{-}{\Stttextmore}}}\Scribtexttt{ }\RktVal{(}\RktVal{(}\RktVal{A}\Scribtexttt{ }\RktVal{\mbox{{-}{\Stttextmore}}}\Scribtexttt{ }\RktVal{B}\RktVal{)}\Scribtexttt{ }\RktVal{\mbox{{-}{\Stttextmore}}}\Scribtexttt{ }\RktVal{(}\RktVal{A}\Scribtexttt{ }\RktVal{\mbox{{-}{\Stttextmore}}}\Scribtexttt{ }\RktVal{C}\RktVal{)}\RktVal{)}\RktVal{)}\RktVal{)}\RktPn{)} \hspace*{\fill}\\
\RktOut{Task is now ({\hbox{\texttt{?}}}0 {\hbox{\texttt{:}}} ((B {-}{\Stttextmore} C) {-}{\Stttextmore} ((A {-}{\Stttextmore} B) {-}{\Stttextmore} (A {-}{\Stttextmore} C))))} \hspace*{\fill}\\
$>$ \RktPn{(}\RktSym{refine}\Scribtexttt{ }\RktVal{0}\Scribtexttt{ }\RktVal{{\textquotesingle}}\RktVal{(}\RktVal{$\lambda$}\Scribtexttt{ }\RktVal{f}\Scribtexttt{ }\RktVal{={\Stttextmore}}\Scribtexttt{ }\RktVal{{\hbox{\texttt{?}}}}\RktVal{)}\RktPn{)} \hspace*{\fill}\\
\RktOut{Task is now (($\lambda$ f ={\Stttextmore} {\hbox{\texttt{?}}}1) {\hbox{\texttt{:}}} ((B {-}{\Stttextmore} C) {-}{\Stttextmore} ((A {-}{\Stttextmore} B) {-}{\Stttextmore} (A {-}{\Stttextmore} C))))} \hspace*{\fill}\\
$>$ \RktPn{(}\RktSym{refine}\Scribtexttt{ }\RktVal{1}\Scribtexttt{ }\RktVal{{\textquotesingle}}\RktVal{(}\RktVal{$\lambda$}\Scribtexttt{ }\RktVal{g}\Scribtexttt{ }\RktVal{={\Stttextmore}}\Scribtexttt{ }\RktVal{{\hbox{\texttt{?}}}}\RktVal{)}\RktPn{)} \hspace*{\fill}\\
\RktOut{Task is now (($\lambda$ f ={\Stttextmore} ($\lambda$ g ={\Stttextmore} {\hbox{\texttt{?}}}2)) {\hbox{\texttt{:}}} ((B {-}{\Stttextmore} C) {-}{\Stttextmore} ((A {-}{\Stttextmore} B) {-}{\Stttextmore} (A {-}{\Stttextmore} C))))} \hspace*{\fill}\\
$>$ \RktPn{(}\RktSym{refine}\Scribtexttt{ }\RktVal{2}\Scribtexttt{ }\RktVal{{\textquotesingle}}\RktVal{(}\RktVal{$\lambda$}\Scribtexttt{ }\RktVal{x}\Scribtexttt{ }\RktVal{={\Stttextmore}}\Scribtexttt{ }\RktVal{{\hbox{\texttt{?}}}}\RktVal{)}\RktPn{)} \hspace*{\fill}\\
\RktOut{Task is now (($\lambda$ f ={\Stttextmore} ($\lambda$ g ={\Stttextmore} ($\lambda$ x ={\Stttextmore} {\hbox{\texttt{?}}}3))) {\hbox{\texttt{:}}} ((B {-}{\Stttextmore} C) {-}{\Stttextmore} ((A {-}{\Stttextmore} B) {-}{\Stttextmore} (A {-}{\Stttextmore} C))))}
$>$ \RktPn{(}\RktSym{refine}\Scribtexttt{ }\RktVal{3}\Scribtexttt{ }\RktVal{{\textquotesingle}}\RktVal{(}\RktVal{f}\Scribtexttt{ }\RktVal{{\hbox{\texttt{?}}}}\RktVal{)}\RktPn{)}\hspace*{\fill}\\
\RktOut{Task is now\hspace*{\fill}\\
(($\lambda$ f ={\Stttextmore} ($\lambda$ g ={\Stttextmore} ($\lambda$ x ={\Stttextmore} (f {\hbox{\texttt{?}}}4)))) {\hbox{\texttt{:}}} ((B {-}{\Stttextmore} C) {-}{\Stttextmore} ((A {-}{\Stttextmore} B) {-}{\Stttextmore} (A {-}{\Stttextmore} C))))}\hspace*{\fill}\\
$>$ \RktPn{(}\RktSym{refine}\Scribtexttt{ }\RktVal{4}\Scribtexttt{ }\RktVal{{\textquotesingle}}\RktVal{(}\RktVal{g}\Scribtexttt{ }\RktVal{{\hbox{\texttt{?}}}}\RktVal{)}\RktPn{)}\hspace*{\fill}\\
\RktOut{Task is now\hspace*{\fill}\\
(($\lambda$ f ={\Stttextmore} ($\lambda$ g ={\Stttextmore} ($\lambda$ x ={\Stttextmore} (f (g {\hbox{\texttt{?}}}5))))) {\hbox{\texttt{:}}} ((B {-}{\Stttextmore} C) {-}{\Stttextmore} ((A {-}{\Stttextmore} B) {-}{\Stttextmore} (A {-}{\Stttextmore} C))))}\hspace*{\fill}\\
$>$ \RktPn{(}\RktSym{refine}\Scribtexttt{ }\RktVal{5}\Scribtexttt{ }\RktVal{{\textquotesingle}}\RktVal{x}\RktPn{)}\hspace*{\fill}\\
\RktOut{Task is now\hspace*{\fill}\\
(($\lambda$ f ={\Stttextmore} ($\lambda$ g ={\Stttextmore} ($\lambda$ x ={\Stttextmore} (f (g x))))) {\hbox{\texttt{:}}} ((B {-}{\Stttextmore} C) {-}{\Stttextmore} ((A {-}{\Stttextmore} B) {-}{\Stttextmore} (A {-}{\Stttextmore} C))))}
}

~

The strength of the metaphor persists for the connectives $\land$ and
$\lor$. A proof of $T \land W$ is a proof of $T$ and a proof of $W$,
so the proof term is a two-field record. We could call it
\RktSym{cons}, but it makes proofs more readable if we call the
constructor (of the proof term that uses the introduction rule)
\RktSym{$\land$-intro}, and call the accessor functions (used in proof
terms that use the two elimination rules) \RktSym{$\land$-elim0}
and \RktSym{$\land$-elim1}.

\begin{mathpar}
\inferrule* [Right=($\land_I$)]{\Gamma\vdash t\La T ~~\Gamma\vdash w\La W}
  {\Gamma\vdash\land_\mathit{intro}~t~w\La T\land W}
 \and
 \inferrule* [Right=($\land_{E0}$)]{\Gamma\vdash v\Ra T\land W}
             {\Gamma\vdash\land_\mathit{elim0}~v\La T}
 \and
 \inferrule* [Right=($\land_{E1}$)]{\Gamma\vdash v\Ra T\land W}
             {\Gamma\vdash\land_\mathit{elim1}~v\La W}
\end{mathpar}

\begin{mathpar}
\inferrule* [Right=($\land_I$)]{\Gamma\vdash t\Ra T ~~\Gamma\vdash w\Ra W}
  {\Gamma\vdash\land_\mathit{intro}~t~w\Ra T\land W}
 \and
 \inferrule* [Right=($\land_{E0}$)]{\Gamma\vdash v\Ra T\land W}
             {\Gamma\vdash\land_\mathit{elim0}~v\Ra T}
 \and
 \inferrule* [Right=($\land_{E1}$)]{\Gamma\vdash v\Ra T\land W}
             {\Gamma\vdash\land_\mathit{elim1}~v\Ra W}
\end{mathpar}

A proof of $T \lor W$ is either a proof of $T$ or a proof of $W$, so
there are two introduction rules and two constructors,\RktSym{$\lor$-intro0}
and \RktSym{$\lor$-intro1}. To use a proof of $T \lor W$, we need a
way to use it if it is a proof of $T$, and if it is a proof of $W$. If
we want to produce a proof of $V$, we must have a proof of $T\ra V$ we
can apply, and a proof of $W \ra V$. The deconstructor
\RktSym{$\lor$-elim} is a variation on a case expression.
(It is normally made a new binding form to avoid mixing connectives in
rules, but we already have a perfectly good binding form, namely lambda.)

\begin{mathpar}
  \inferrule* [Right=($\lor_{I0}$)]{\Gamma\vdash t\La T}
              {\Gamma\vdash\lor_\mathit{intro0}~t\La T\lor W}
\and
\inferrule* [Right=($\lor_{I1}$)]{\Gamma\vdash w\La W}
            {\Gamma\vdash\lor_\mathit{intro1}~w\La T\lor W}
\and
\inferrule* [Right=($\lor_E$)]
            {\Gamma\vdash v\Ra T\lor W \\\\ \Gamma\vdash f\La T\ra V
              ~~~~\Gamma\vdash g\La W\ra V}
  {\Gamma\vdash\lor_\mathit{elim}~v~f~g\La V}
\end{mathpar}

Finally, we have negation, whose treatment remains intuitive but may
seem incomplete. The logical constant $\bot$ denotes an absurdity or
contradiction (we use this for both the term and its type).
$\bot$ has no constructor, but its eliminator \RktSym{$\bot$-elim}
can have any type. We define $\lnot T$ to be $T \ra \bot$, and use this as a
desugaring rule in the parser.

\begin{mathpar}
  \inferrule* [Right=($\bot_{E}$)]{\Gamma\vdash t\La\bot}
              {\Gamma\vdash \bot_\mathit{elim}~t\La T}
\end{mathpar}

There are now more statements we cannot prove: $A \lor \lnot A$,
$\lnot \lnot A \ra A$, and some of deMorgan's laws. But there are
plenty of things we can prove, and this is a good point to assign some
proof exercises -- once students implement the necessary changes to
the parser, pretty-printer, type checking, and type inference
functions, because these are also reasonable programming exercises.
Each connective requires a very small number of additional lines of
code. 

While we are writing proof terms that are functions in a
language tantalizingly close to the one we are programming in, there
is no notion of computation in that language so far. We draw
conclusions about the AST, but we do not interpret or otherwise
execute the proof-programs. In effect, we have only an evocative tree
representation using S-expressions. This will change in the next
section.

\section{Predicate Logic}\label{Pred}

A conventional treatment introduces predicate logic by adding
relational symbols and first-order quantification to propositional
logic, possibly followed by axioms for equality and arithmetic.  But
just as students benefit from the use of higher-order functions in
programming, they can benefit from the increased expressivity of
higher-order logic. Our goal in this section, and in the development
of Proust, is to reach the point where we can state and prove a
suitable translation of ``For all natural numbers $n$,
$\mathit{plus}(n,0)=n$'', where $\mathit{plus}$ is a user-implemented
function using structural induction on the first argument (necessitating a
proof by induction).

\clearpage

In the previous section, we maintained a grammatical distinction
between terms and types. But this example illustrates the need to
eliminate this distinction. We rewind to the first program of
section \ref{Pred},
handling the implicational fragment of propositional logic without
holes, and merge the grammar rules.

~

\label{t:x28part_x28gentag_0x29x29}

\begin{bigtabular}{@{\bigtableleftpad}r@{}l@{}c@{}l@{}l@{}}
\hbox{\mbox{\hphantom{\Scribtexttt{xx}}}\RktVar{expr}} &
\hbox{\Scribtexttt{~}} &
\hbox{=} &
\hbox{\Scribtexttt{~}} &
\hbox{\RktPn{(}\RktVar{$\lambda$}\mbox{\hphantom{\Scribtexttt{x}}}\RktVar{x}\mbox{\hphantom{\Scribtexttt{x}}}\RktVar{={\Stttextmore}}\mbox{\hphantom{\Scribtexttt{x}}}\RktVar{expr}\RktPn{)}} \\
\hbox{\Scribtexttt{~}} &
\hbox{\Scribtexttt{~}} &
\hbox{$|$} &
\hbox{\Scribtexttt{~}} &
\hbox{\RktPn{(}\RktVar{expr}\mbox{\hphantom{\Scribtexttt{x}}}\RktVar{expr}\RktPn{)}} \\
\hbox{\Scribtexttt{~}} &
\hbox{\Scribtexttt{~}} &
\hbox{$|$} &
\hbox{\Scribtexttt{~}} &
\hbox{\RktPn{(}\RktVar{expr}\mbox{\hphantom{\Scribtexttt{x}}}\RktVar{{\hbox{\texttt{:}}}}\mbox{\hphantom{\Scribtexttt{x}}}\RktVar{expr}\RktPn{)}} \\
\hbox{\Scribtexttt{~}} &
\hbox{\Scribtexttt{~}} &
\hbox{$|$} &
\hbox{\Scribtexttt{~}} &
\hbox{\RktVar{x}} \\
\hbox{\Scribtexttt{~}} &
\hbox{\Scribtexttt{~}} &
\hbox{$|$} &
\hbox{\Scribtexttt{~}} &
\hbox{\RktPn{(}\RktVar{expr}\mbox{\hphantom{\Scribtexttt{x}}}\RktVar{{-}{\Stttextmore}}\mbox{\hphantom{\Scribtexttt{x}}}\RktVar{expr}\RktPn{)}} \\
\hbox{\Scribtexttt{~}} &
\hbox{\Scribtexttt{~}} &
\hbox{$|$} &
\hbox{\Scribtexttt{~}} &
\hbox{\RktVar{X}}\end{bigtabular}

~

The grammar is too permissive. We enforce well-formedness in code,
adding a \RktSym{Type} structure which will be the type inferred for
a term which satisfies the previous grammar rule for types. We will
soon add $\mathit{Type}$ to the term language, but not yet, as we need
to refactor the program so that the structure of our functions mirrors
that of the merged grammar. Having done that, we are ready to consider
quantification. As before, we will reason about what the proof rules
should be, and what proof terms should look like.

We will write a for-all statement as $\forall (x:T) \ra W$, where $T$
and $W$ are terms (Agda uses a similar syntax).
How do we use such a statement in an informal proof?
We instantiate $x$ with a specific value $t$ of type $T$,
substituting $t$ for $x$ everywhere that $x$ occurs in $W$. Using the
notation $W[x\mapsto t]$ for this substitution, and leaving the proof
terms unspecified for the moment, the rule looks like this:

\begin{mathpar}
\inferrule* [Right=($\forall_{E}$)]{\Gamma\vdash ?\Ra\forall
  (x:T)\ra W \\ \Gamma\vdash t\La T}
            {\Gamma\vdash ?\La W[x\mapsto t]}
\end{mathpar}

If we consider the special case where $W$ does not use $x$, the
substitution has no effect, and the rule looks like implication
elimination. This suggests that for-all is a generalization of
implication, that the introduction proof term should be a
generalization of lambda, and that the elimination proof term
should be function application. We can keep implication in
the term language for convenience, and desugar it to an
equivalent for-all. This version of for-all is a dependent product type, often
written $\Pi$ in the literature (as in the title of \cite{ADLO}).

\begin{mathpar}
  \inferrule* [Right=($\forall_{I}$)]
              {\Gamma,x:T\vdash t\La W}
              {\Gamma\vdash\la x.t\La\forall (x:T)\ra W}
 \and
 \inferrule* [Right=($\forall_{E}$)]
             {\Gamma\vdash f\Ra\forall (x:T)\ra W \\ \Gamma\vdash t\La
               T}
             {\Gamma\vdash f~t\La W[x\mapsto t]}
\end{mathpar}

Once we add $\mathit{Type}$ as a constant to the language of terms, we
can, for example, write the polymorphic identity function as $\la x.x
: \forall (y:\mathit{Type}) \ra (y \ra y)$. We remove propositional
variables from the term language; we only have variables that are
parameters. But if $\mathit{Type}$ is a value, what is its type? Here
I make a decision that is contrary to our goal of a reliable proof
assistant: $\mathit{Type}$ has type $\mathit{Type}$. This was done by
Martin-L\"of in the earliest version of his type theory, but it was
shown by Girard \cite{Gir} to result in an inconsistent logic. The
proof is complicated and not a simple, direct construction as in
Russell's paradox, so it is unlikely to be an issue for students.
Inconsistency can be avoided by a hierarchy of types
($\mathit{Type}_0:\mathit{Type}_1$, etc.),
and this is what is done in Coq and Agda,
though Coq hides the details from the user until they become
relevant. Managing this hierarchy is not difficult, but it is tedious,
and some of the tutorial materials discussed in section \ref{Intro}
choose to avoid it in the same fashion as I have
(\cite{ADLO},\cite{LMS},\cite{Weir}), while
others tackle it (\cite{Bauer}). Avoidance seems best in a context
where we are moving towards use of full-featured assistants.

Implementing the above rules requires implementing substitution, with
attention to variable reuse and variable capture. There are places in
the code where the Racket predicate \RktSym{equal?} is used to compare
types for structural equality (for example, in the implementation of
the ``turn'' rule). But the polymorphic identity function mentioned
above could equally well be typed as $\forall (z:\mathit{Type}) \ra (z
\ra z)$. The name of the variable should not matter. In other words,
we need alpha-equivalence. These topics are typically covered in a
full treatment of the lambda calculus (and in a principled treatment
of predicate logic), but students in L\&C will not have seen them. 
Since one goal is for them to understand the implementation,
they should be covered in lecture.

As our proof terms are going to get more complicated, we add an
association list of global definitions, managed by functions such as
\RktSym{def}, which typechecks an annotated term and adds it to the
list with a symbolic name. But substituting such a definition for the
use of a name will result in expressions that require simplification
through reduction. We have already coded the basic substitution
mechanism. Reduction to weak normal head form suffices in some places,
but strong reduction (full beta-reduction) is needed in
others. Definitional equality is the alpha-equivalence of two
strongly-reduced expressions.

All of this takes considerably more care to present in the classroom
than I have taken here.  A proper presentation takes students through
the reasons for each additional requirement, by pointing out
expressions that we would expect to typecheck but do not in an
incomplete implementation.  The code base is still under two hundred lines (an
additional 10 lines for alpha equivalence, 30 for substitution, 30 for
reduction and equivalence, 20 for support for definitions).

The reward for all this work is a surprisingly expressive language,
considering its size. To start with, we can implement Church encodings.
The Boolean type is implemented by its eliminator, namely
if-then-else, which makes it easy to implement other Boolean functions.

~

\label{t:x28part_x28gentag_0x29x29}

\begin{RktBlk}\begin{SingleColumn}\RktPn{(}\RktSym{def}\mbox{\hphantom{\Scribtexttt{x}}}\RktVal{{\textquotesingle}}\RktVal{bool}\mbox{\hphantom{\Scribtexttt{x}}}\RktVal{{\textquotesingle}}\RktVal{(}\RktVal{(}\RktVal{$\forall$}\mbox{\hphantom{\Scribtexttt{x}}}\RktVal{(}\RktVal{x}\mbox{\hphantom{\Scribtexttt{x}}}\RktVal{{\hbox{\texttt{:}}}}\mbox{\hphantom{\Scribtexttt{x}}}\RktVal{Type}\RktVal{)}\mbox{\hphantom{\Scribtexttt{x}}}\RktVal{\mbox{{-}{\Stttextmore}}}\mbox{\hphantom{\Scribtexttt{x}}}\RktVal{(}\RktVal{x}\mbox{\hphantom{\Scribtexttt{x}}}\RktVal{\mbox{{-}{\Stttextmore}}}\mbox{\hphantom{\Scribtexttt{x}}}\RktVal{(}\RktVal{x}\mbox{\hphantom{\Scribtexttt{x}}}\RktVal{\mbox{{-}{\Stttextmore}}}\mbox{\hphantom{\Scribtexttt{x}}}\RktVal{x}\RktVal{)}\RktVal{)}\RktVal{)}\mbox{\hphantom{\Scribtexttt{x}}}\RktVal{{\hbox{\texttt{:}}}}\mbox{\hphantom{\Scribtexttt{x}}}\RktVal{Type}\RktVal{)}\RktPn{)}

\RktPn{(}\RktSym{def}\mbox{\hphantom{\Scribtexttt{x}}}\RktVal{{\textquotesingle}}\RktVal{true}\mbox{\hphantom{\Scribtexttt{x}}}\RktVal{{\textquotesingle}}\RktVal{(}\RktVal{(}\RktVal{$\lambda$}\mbox{\hphantom{\Scribtexttt{x}}}\RktVal{x}\mbox{\hphantom{\Scribtexttt{x}}}\RktVal{={\Stttextmore}}\mbox{\hphantom{\Scribtexttt{x}}}\RktVal{(}\RktVal{$\lambda$}\mbox{\hphantom{\Scribtexttt{x}}}\RktVal{y}\mbox{\hphantom{\Scribtexttt{x}}}\RktVal{={\Stttextmore}}\mbox{\hphantom{\Scribtexttt{x}}}\RktVal{(}\RktVal{$\lambda$}\mbox{\hphantom{\Scribtexttt{x}}}\RktVal{z}\mbox{\hphantom{\Scribtexttt{x}}}\RktVal{={\Stttextmore}}\mbox{\hphantom{\Scribtexttt{x}}}\RktVal{y}\RktVal{)}\RktVal{)}\RktVal{)}\mbox{\hphantom{\Scribtexttt{x}}}\RktVal{{\hbox{\texttt{:}}}}\mbox{\hphantom{\Scribtexttt{x}}}\RktVal{bool}\RktVal{)}\RktPn{)}

\RktPn{(}\RktSym{def}\mbox{\hphantom{\Scribtexttt{x}}}\RktVal{{\textquotesingle}}\RktVal{false}\mbox{\hphantom{\Scribtexttt{x}}}\RktVal{{\textquotesingle}}\RktVal{(}\RktVal{(}\RktVal{$\lambda$}\mbox{\hphantom{\Scribtexttt{x}}}\RktVal{x}\mbox{\hphantom{\Scribtexttt{x}}}\RktVal{={\Stttextmore}}\mbox{\hphantom{\Scribtexttt{x}}}\RktVal{(}\RktVal{$\lambda$}\mbox{\hphantom{\Scribtexttt{x}}}\RktVal{y}\mbox{\hphantom{\Scribtexttt{x}}}\RktVal{={\Stttextmore}}\mbox{\hphantom{\Scribtexttt{x}}}\RktVal{(}\RktVal{$\lambda$}\mbox{\hphantom{\Scribtexttt{x}}}\RktVal{z}\mbox{\hphantom{\Scribtexttt{x}}}\RktVal{={\Stttextmore}}\mbox{\hphantom{\Scribtexttt{x}}}\RktVal{z}\RktVal{)}\RktVal{)}\RktVal{)}\mbox{\hphantom{\Scribtexttt{x}}}\RktVal{{\hbox{\texttt{:}}}}\mbox{\hphantom{\Scribtexttt{x}}}\RktVal{bool}\RktVal{)}\RktPn{)}

\RktPn{(}\RktSym{def}\mbox{\hphantom{\Scribtexttt{x}}}\RktVal{{\textquotesingle}}\RktVal{band}\mbox{\hphantom{\Scribtexttt{xx}}}\RktVal{{\textquotesingle}}\RktVal{(}\RktVal{(}\RktVal{(}\RktVal{$\lambda$}\mbox{\hphantom{\Scribtexttt{x}}}\RktVal{x}\mbox{\hphantom{\Scribtexttt{x}}}\RktVal{={\Stttextmore}}\mbox{\hphantom{\Scribtexttt{x}}}\RktVal{(}\RktVal{$\lambda$}\mbox{\hphantom{\Scribtexttt{x}}}\RktVal{y}\mbox{\hphantom{\Scribtexttt{x}}}\RktVal{={\Stttextmore}}\mbox{\hphantom{\Scribtexttt{x}}}\RktVal{(}\RktVal{(}\RktVal{(}\RktVal{x}\mbox{\hphantom{\Scribtexttt{x}}}\RktVal{bool}\RktVal{)}\mbox{\hphantom{\Scribtexttt{x}}}\RktVal{y}\RktVal{)}\mbox{\hphantom{\Scribtexttt{x}}}\RktVal{false}\RktVal{)}\RktVal{)}\RktVal{)}\RktVal{)}

\mbox{\hphantom{\Scribtexttt{xxxxx}}}\RktVal{{\hbox{\texttt{:}}}}\mbox{\hphantom{\Scribtexttt{x}}}\RktVal{(}\RktVal{bool}\mbox{\hphantom{\Scribtexttt{x}}}\RktVal{\mbox{{-}{\Stttextmore}}}\mbox{\hphantom{\Scribtexttt{x}}}\RktVal{(}\RktVal{bool}\mbox{\hphantom{\Scribtexttt{x}}}\RktVal{\mbox{{-}{\Stttextmore}}}\mbox{\hphantom{\Scribtexttt{x}}}\RktVal{bool}\RktVal{)}\RktVal{)}\RktVal{)}\RktPn{)}\end{SingleColumn}\end{RktBlk}

~

We can implement logical AND ($\land$), and show that it
commutes. Logical OR ($\lor$) is also possible.

~

\label{t:x28part_x28gentag_0x29x29}

\begin{RktBlk}\begin{SingleColumn}\RktPn{(}\RktSym{def}\mbox{\hphantom{\Scribtexttt{x}}}\RktVal{{\textquotesingle}}\RktVal{and}\mbox{\hphantom{\Scribtexttt{x}}}\RktVal{{\textquotesingle}}\RktVal{(}\RktVal{(}\RktVal{$\lambda$}\mbox{\hphantom{\Scribtexttt{x}}}\RktVal{p}\mbox{\hphantom{\Scribtexttt{x}}}\RktVal{={\Stttextmore}}\mbox{\hphantom{\Scribtexttt{x}}}\RktVal{(}\RktVal{$\lambda$}\mbox{\hphantom{\Scribtexttt{x}}}\RktVal{q}\mbox{\hphantom{\Scribtexttt{x}}}\RktVal{={\Stttextmore}}\mbox{\hphantom{\Scribtexttt{x}}}\RktVal{(}\RktVal{$\forall$}\mbox{\hphantom{\Scribtexttt{x}}}\RktVal{(}\RktVal{c}\mbox{\hphantom{\Scribtexttt{x}}}\RktVal{{\hbox{\texttt{:}}}}\mbox{\hphantom{\Scribtexttt{x}}}\RktVal{Type}\RktVal{)}\mbox{\hphantom{\Scribtexttt{x}}}\RktVal{\mbox{{-}{\Stttextmore}}}\mbox{\hphantom{\Scribtexttt{x}}}\RktVal{(}\RktVal{(}\RktVal{p}\mbox{\hphantom{\Scribtexttt{x}}}\RktVal{\mbox{{-}{\Stttextmore}}}\mbox{\hphantom{\Scribtexttt{x}}}\RktVal{(}\RktVal{q}\mbox{\hphantom{\Scribtexttt{x}}}\RktVal{\mbox{{-}{\Stttextmore}}}\mbox{\hphantom{\Scribtexttt{x}}}\RktVal{c}\RktVal{)}\RktVal{)}\mbox{\hphantom{\Scribtexttt{x}}}\RktVal{\mbox{{-}{\Stttextmore}}}\mbox{\hphantom{\Scribtexttt{x}}}\RktVal{c}\RktVal{)}\RktVal{)}\RktVal{)}\RktVal{)}

\mbox{\hphantom{\Scribtexttt{xxxxxxxxxxxxx}}}\RktVal{{\hbox{\texttt{:}}}}\mbox{\hphantom{\Scribtexttt{x}}}\RktVal{(}\RktVal{Type}\mbox{\hphantom{\Scribtexttt{x}}}\RktVal{\mbox{{-}{\Stttextmore}}}\mbox{\hphantom{\Scribtexttt{x}}}\RktVal{(}\RktVal{Type}\mbox{\hphantom{\Scribtexttt{x}}}\RktVal{\mbox{{-}{\Stttextmore}}}\mbox{\hphantom{\Scribtexttt{x}}}\RktVal{Type}\RktVal{)}\RktVal{)}\RktVal{)}\RktPn{)}

\RktPn{(}\RktSym{def}\mbox{\hphantom{\Scribtexttt{x}}}\RktVal{{\textquotesingle}}\RktVal{conj}\mbox{\hphantom{\Scribtexttt{x}}}\RktVal{{\textquotesingle}}\RktVal{(}\RktVal{(}\RktVal{$\lambda$}\mbox{\hphantom{\Scribtexttt{x}}}\RktVal{p}\mbox{\hphantom{\Scribtexttt{x}}}\RktVal{={\Stttextmore}}\mbox{\hphantom{\Scribtexttt{x}}}\RktVal{(}\RktVal{$\lambda$}\mbox{\hphantom{\Scribtexttt{x}}}\RktVal{q}\mbox{\hphantom{\Scribtexttt{x}}}\RktVal{={\Stttextmore}}\mbox{\hphantom{\Scribtexttt{x}}}\RktVal{(}\RktVal{$\lambda$}\mbox{\hphantom{\Scribtexttt{x}}}\RktVal{x}\mbox{\hphantom{\Scribtexttt{x}}}\RktVal{={\Stttextmore}}\mbox{\hphantom{\Scribtexttt{x}}}\RktVal{(}\RktVal{$\lambda$}\mbox{\hphantom{\Scribtexttt{x}}}\RktVal{y}\mbox{\hphantom{\Scribtexttt{x}}}\RktVal{={\Stttextmore}}\mbox{\hphantom{\Scribtexttt{x}}}\RktVal{(}\RktVal{$\lambda$}\mbox{\hphantom{\Scribtexttt{x}}}\RktVal{c}\mbox{\hphantom{\Scribtexttt{x}}}\RktVal{={\Stttextmore}}\mbox{\hphantom{\Scribtexttt{x}}}\RktVal{(}\RktVal{$\lambda$}\mbox{\hphantom{\Scribtexttt{x}}}\RktVal{f}\mbox{\hphantom{\Scribtexttt{x}}}\RktVal{={\Stttextmore}}\mbox{\hphantom{\Scribtexttt{x}}}\RktVal{(}\RktVal{(}\RktVal{f}\mbox{\hphantom{\Scribtexttt{x}}}\RktVal{x}\RktVal{)}\mbox{\hphantom{\Scribtexttt{x}}}\RktVal{y}\RktVal{)}\RktVal{)}\RktVal{)}\RktVal{)}\RktVal{)}\RktVal{)}\RktVal{)}

\mbox{\hphantom{\Scribtexttt{xxxxxxxxxxxxx}}}\RktVal{{\hbox{\texttt{:}}}}\mbox{\hphantom{\Scribtexttt{x}}}\RktVal{(}\RktVal{$\forall$}\mbox{\hphantom{\Scribtexttt{x}}}\RktVal{(}\RktVal{p}\mbox{\hphantom{\Scribtexttt{x}}}\RktVal{{\hbox{\texttt{:}}}}\mbox{\hphantom{\Scribtexttt{x}}}\RktVal{Type}\RktVal{)}\mbox{\hphantom{\Scribtexttt{x}}}\RktVal{\mbox{{-}{\Stttextmore}}}\mbox{\hphantom{\Scribtexttt{x}}}\RktVal{(}\RktVal{$\forall$}\mbox{\hphantom{\Scribtexttt{x}}}\RktVal{(}\RktVal{q}\mbox{\hphantom{\Scribtexttt{x}}}\RktVal{{\hbox{\texttt{:}}}}\mbox{\hphantom{\Scribtexttt{x}}}\RktVal{Type}\RktVal{)}\mbox{\hphantom{\Scribtexttt{x}}}\RktVal{\mbox{{-}{\Stttextmore}}}\mbox{\hphantom{\Scribtexttt{x}}}\RktVal{(}\RktVal{p}\mbox{\hphantom{\Scribtexttt{x}}}\RktVal{\mbox{{-}{\Stttextmore}}}\mbox{\hphantom{\Scribtexttt{x}}}\RktVal{(}\RktVal{q}\mbox{\hphantom{\Scribtexttt{x}}}\RktVal{\mbox{{-}{\Stttextmore}}}\mbox{\hphantom{\Scribtexttt{x}}}\RktVal{(}\RktVal{(}\RktVal{and}\mbox{\hphantom{\Scribtexttt{x}}}\RktVal{p}\RktVal{)}\mbox{\hphantom{\Scribtexttt{x}}}\RktVal{q}\RktVal{)}\RktVal{)}\RktVal{)}\RktVal{)}\RktVal{)}\RktVal{)}\RktPn{)}

\RktPn{(}\RktSym{def}\mbox{\hphantom{\Scribtexttt{x}}}\RktVal{{\textquotesingle}}\RktVal{proj1}\mbox{\hphantom{\Scribtexttt{x}}}\RktVal{{\textquotesingle}}\RktVal{(}\RktVal{(}\RktVal{$\lambda$}\mbox{\hphantom{\Scribtexttt{x}}}\RktVal{p}\mbox{\hphantom{\Scribtexttt{x}}}\RktVal{={\Stttextmore}}\mbox{\hphantom{\Scribtexttt{x}}}\RktVal{(}\RktVal{$\lambda$}\mbox{\hphantom{\Scribtexttt{x}}}\RktVal{q}\mbox{\hphantom{\Scribtexttt{x}}}\RktVal{={\Stttextmore}}\mbox{\hphantom{\Scribtexttt{x}}}\RktVal{(}\RktVal{$\lambda$}\mbox{\hphantom{\Scribtexttt{x}}}\RktVal{a}\mbox{\hphantom{\Scribtexttt{x}}}\RktVal{={\Stttextmore}}\mbox{\hphantom{\Scribtexttt{x}}}\RktVal{(}\RktVal{(}\RktVal{a}\mbox{\hphantom{\Scribtexttt{x}}}\RktVal{p}\RktVal{)}\mbox{\hphantom{\Scribtexttt{x}}}\RktVal{(}\RktVal{$\lambda$}\mbox{\hphantom{\Scribtexttt{x}}}\RktVal{x}\mbox{\hphantom{\Scribtexttt{x}}}\RktVal{={\Stttextmore}}\mbox{\hphantom{\Scribtexttt{x}}}\RktVal{(}\RktVal{$\lambda$}\mbox{\hphantom{\Scribtexttt{x}}}\RktVal{y}\mbox{\hphantom{\Scribtexttt{x}}}\RktVal{={\Stttextmore}}\mbox{\hphantom{\Scribtexttt{x}}}\RktVal{x}\RktVal{)}\RktVal{)}\RktVal{)}\RktVal{)}\RktVal{)}\RktVal{)}

\mbox{\hphantom{\Scribtexttt{xxxxxxxxxxxxxx}}}\RktVal{{\hbox{\texttt{:}}}}\mbox{\hphantom{\Scribtexttt{x}}}\RktVal{(}\RktVal{$\forall$}\mbox{\hphantom{\Scribtexttt{x}}}\RktVal{(}\RktVal{p}\mbox{\hphantom{\Scribtexttt{x}}}\RktVal{{\hbox{\texttt{:}}}}\mbox{\hphantom{\Scribtexttt{x}}}\RktVal{Type}\RktVal{)}\mbox{\hphantom{\Scribtexttt{x}}}\RktVal{\mbox{{-}{\Stttextmore}}}\mbox{\hphantom{\Scribtexttt{x}}}\RktVal{(}\RktVal{$\forall$}\mbox{\hphantom{\Scribtexttt{x}}}\RktVal{(}\RktVal{q}\mbox{\hphantom{\Scribtexttt{x}}}\RktVal{{\hbox{\texttt{:}}}}\mbox{\hphantom{\Scribtexttt{x}}}\RktVal{Type}\RktVal{)}\mbox{\hphantom{\Scribtexttt{x}}}\RktVal{\mbox{{-}{\Stttextmore}}}\mbox{\hphantom{\Scribtexttt{x}}}\RktVal{(}\RktVal{(}\RktVal{(}\RktVal{and}\mbox{\hphantom{\Scribtexttt{x}}}\RktVal{p}\RktVal{)}\mbox{\hphantom{\Scribtexttt{x}}}\RktVal{q}\RktVal{)}\mbox{\hphantom{\Scribtexttt{x}}}\RktVal{\mbox{{-}{\Stttextmore}}}\mbox{\hphantom{\Scribtexttt{x}}}\RktVal{p}\RktVal{)}\RktVal{)}\RktVal{)}\RktVal{)}\RktPn{)}

\RktPn{(}\RktSym{def}\mbox{\hphantom{\Scribtexttt{x}}}\RktVal{{\textquotesingle}}\RktVal{proj2}\mbox{\hphantom{\Scribtexttt{x}}}\RktVal{{\textquotesingle}}\RktVal{(}\RktVal{(}\RktVal{$\lambda$}\mbox{\hphantom{\Scribtexttt{x}}}\RktVal{p}\mbox{\hphantom{\Scribtexttt{x}}}\RktVal{={\Stttextmore}}\mbox{\hphantom{\Scribtexttt{x}}}\RktVal{(}\RktVal{$\lambda$}\mbox{\hphantom{\Scribtexttt{x}}}\RktVal{q}\mbox{\hphantom{\Scribtexttt{x}}}\RktVal{={\Stttextmore}}\mbox{\hphantom{\Scribtexttt{x}}}\RktVal{(}\RktVal{$\lambda$}\mbox{\hphantom{\Scribtexttt{x}}}\RktVal{a}\mbox{\hphantom{\Scribtexttt{x}}}\RktVal{={\Stttextmore}}\mbox{\hphantom{\Scribtexttt{x}}}\RktVal{(}\RktVal{(}\RktVal{a}\mbox{\hphantom{\Scribtexttt{x}}}\RktVal{q}\RktVal{)}\mbox{\hphantom{\Scribtexttt{x}}}\RktVal{(}\RktVal{$\lambda$}\mbox{\hphantom{\Scribtexttt{x}}}\RktVal{x}\mbox{\hphantom{\Scribtexttt{x}}}\RktVal{={\Stttextmore}}\mbox{\hphantom{\Scribtexttt{x}}}\RktVal{(}\RktVal{$\lambda$}\mbox{\hphantom{\Scribtexttt{x}}}\RktVal{y}\mbox{\hphantom{\Scribtexttt{x}}}\RktVal{={\Stttextmore}}\mbox{\hphantom{\Scribtexttt{x}}}\RktVal{y}\RktVal{)}\RktVal{)}\RktVal{)}\RktVal{)}\RktVal{)}\RktVal{)}

\mbox{\hphantom{\Scribtexttt{xxxxxxxxxxxxxx}}}\RktVal{{\hbox{\texttt{:}}}}\mbox{\hphantom{\Scribtexttt{x}}}\RktVal{(}\RktVal{$\forall$}\mbox{\hphantom{\Scribtexttt{x}}}\RktVal{(}\RktVal{p}\mbox{\hphantom{\Scribtexttt{x}}}\RktVal{{\hbox{\texttt{:}}}}\mbox{\hphantom{\Scribtexttt{x}}}\RktVal{Type}\RktVal{)}\mbox{\hphantom{\Scribtexttt{x}}}\RktVal{\mbox{{-}{\Stttextmore}}}\mbox{\hphantom{\Scribtexttt{x}}}\RktVal{(}\RktVal{$\forall$}\mbox{\hphantom{\Scribtexttt{x}}}\RktVal{(}\RktVal{q}\mbox{\hphantom{\Scribtexttt{x}}}\RktVal{{\hbox{\texttt{:}}}}\mbox{\hphantom{\Scribtexttt{x}}}\RktVal{Type}\RktVal{)}\mbox{\hphantom{\Scribtexttt{x}}}\RktVal{\mbox{{-}{\Stttextmore}}}\mbox{\hphantom{\Scribtexttt{x}}}\RktVal{(}\RktVal{(}\RktVal{(}\RktVal{and}\mbox{\hphantom{\Scribtexttt{x}}}\RktVal{p}\RktVal{)}\mbox{\hphantom{\Scribtexttt{x}}}\RktVal{q}\RktVal{)}\mbox{\hphantom{\Scribtexttt{x}}}\RktVal{\mbox{{-}{\Stttextmore}}}\mbox{\hphantom{\Scribtexttt{x}}}\RktVal{q}\RktVal{)}\RktVal{)}\RktVal{)}\RktVal{)}\RktPn{)}

\RktPn{(}\RktSym{def}\mbox{\hphantom{\Scribtexttt{x}}}\RktVal{{\textquotesingle}}\RktVal{and{-}commutes}

\mbox{\hphantom{\Scribtexttt{xx}}}\RktVal{{\textquotesingle}}\RktVal{(}\RktVal{(}\RktVal{$\lambda$}\mbox{\hphantom{\Scribtexttt{x}}}\RktVal{p}\mbox{\hphantom{\Scribtexttt{x}}}\RktVal{={\Stttextmore}}\mbox{\hphantom{\Scribtexttt{x}}}\RktVal{(}\RktVal{$\lambda$}\mbox{\hphantom{\Scribtexttt{x}}}\RktVal{q}\mbox{\hphantom{\Scribtexttt{x}}}\RktVal{={\Stttextmore}}\mbox{\hphantom{\Scribtexttt{x}}}\RktVal{(}\RktVal{$\lambda$}\mbox{\hphantom{\Scribtexttt{x}}}\RktVal{a}\mbox{\hphantom{\Scribtexttt{x}}}\RktVal{={\Stttextmore}}\mbox{\hphantom{\Scribtexttt{x}}}\RktVal{(}\RktVal{(}\RktVal{(}\RktVal{(}\RktVal{conj}\mbox{\hphantom{\Scribtexttt{x}}}\RktVal{q}\RktVal{)}\mbox{\hphantom{\Scribtexttt{x}}}\RktVal{p}\RktVal{)}\mbox{\hphantom{\Scribtexttt{x}}}\RktVal{(}\RktVal{(}\RktVal{(}\RktVal{proj2}\mbox{\hphantom{\Scribtexttt{x}}}\RktVal{p}\RktVal{)}\mbox{\hphantom{\Scribtexttt{x}}}\RktVal{q}\RktVal{)}\mbox{\hphantom{\Scribtexttt{x}}}\RktVal{a}\RktVal{)}\RktVal{)}\mbox{\hphantom{\Scribtexttt{x}}}\RktVal{(}\RktVal{(}\RktVal{(}\RktVal{proj1}\mbox{\hphantom{\Scribtexttt{x}}}\RktVal{p}\RktVal{)}\mbox{\hphantom{\Scribtexttt{x}}}\RktVal{q}\RktVal{)}\mbox{\hphantom{\Scribtexttt{x}}}\RktVal{a}\RktVal{)}\RktVal{)}\RktVal{)}\RktVal{)}\RktVal{)}

\mbox{\hphantom{\Scribtexttt{xxxxx}}}\RktVal{{\hbox{\texttt{:}}}}\mbox{\hphantom{\Scribtexttt{x}}}\RktVal{(}\RktVal{$\forall$}\mbox{\hphantom{\Scribtexttt{x}}}\RktVal{(}\RktVal{p}\mbox{\hphantom{\Scribtexttt{x}}}\RktVal{{\hbox{\texttt{:}}}}\mbox{\hphantom{\Scribtexttt{x}}}\RktVal{Type}\RktVal{)}\mbox{\hphantom{\Scribtexttt{x}}}\RktVal{\mbox{{-}{\Stttextmore}}}\mbox{\hphantom{\Scribtexttt{x}}}\RktVal{(}\RktVal{$\forall$}\mbox{\hphantom{\Scribtexttt{x}}}\RktVal{(}\RktVal{q}\mbox{\hphantom{\Scribtexttt{x}}}\RktVal{{\hbox{\texttt{:}}}}\mbox{\hphantom{\Scribtexttt{x}}}\RktVal{Type}\RktVal{)}\mbox{\hphantom{\Scribtexttt{x}}}\RktVal{\mbox{{-}{\Stttextmore}}}\mbox{\hphantom{\Scribtexttt{x}}}\RktVal{(}\RktVal{(}\RktVal{(}\RktVal{and}\mbox{\hphantom{\Scribtexttt{x}}}\RktVal{p}\RktVal{)}\mbox{\hphantom{\Scribtexttt{x}}}\RktVal{q}\RktVal{)}\mbox{\hphantom{\Scribtexttt{x}}}\RktVal{\mbox{{-}{\Stttextmore}}}\mbox{\hphantom{\Scribtexttt{x}}}\RktVal{(}\RktVal{(}\RktVal{and}\mbox{\hphantom{\Scribtexttt{x}}}\RktVal{q}\RktVal{)}\mbox{\hphantom{\Scribtexttt{x}}}\RktVal{p}\RktVal{)}\RktVal{)}\RktVal{)}\RktVal{)}\RktVal{)}\RktPn{)}\end{SingleColumn}\end{RktBlk}

\clearpage

Arithmetic can be implemented using Church numerals.

~

\label{t:x28part_x28gentag_0x29x29}

\begin{RktBlk}\begin{SingleColumn}\RktPn{(}\RktSym{def}\mbox{\hphantom{\Scribtexttt{x}}}\RktVal{{\textquotesingle}}\RktVal{nat}\mbox{\hphantom{\Scribtexttt{x}}}\RktVal{{\textquotesingle}}\RktVal{(}\RktVal{(}\RktVal{$\forall$}\mbox{\hphantom{\Scribtexttt{x}}}\RktVal{(}\RktVal{x}\mbox{\hphantom{\Scribtexttt{x}}}\RktVal{{\hbox{\texttt{:}}}}\mbox{\hphantom{\Scribtexttt{x}}}\RktVal{Type}\RktVal{)}\mbox{\hphantom{\Scribtexttt{x}}}\RktVal{\mbox{{-}{\Stttextmore}}}\mbox{\hphantom{\Scribtexttt{x}}}\RktVal{(}\RktVal{x}\mbox{\hphantom{\Scribtexttt{x}}}\RktVal{\mbox{{-}{\Stttextmore}}}\mbox{\hphantom{\Scribtexttt{x}}}\RktVal{(}\RktVal{(}\RktVal{x}\mbox{\hphantom{\Scribtexttt{x}}}\RktVal{\mbox{{-}{\Stttextmore}}}\mbox{\hphantom{\Scribtexttt{x}}}\RktVal{x}\RktVal{)}\mbox{\hphantom{\Scribtexttt{x}}}\RktVal{\mbox{{-}{\Stttextmore}}}\mbox{\hphantom{\Scribtexttt{x}}}\RktVal{x}\RktVal{)}\RktVal{)}\RktVal{)}\mbox{\hphantom{\Scribtexttt{x}}}\RktVal{{\hbox{\texttt{:}}}}\mbox{\hphantom{\Scribtexttt{x}}}\RktVal{Type}\RktVal{)}\RktPn{)}

\RktPn{(}\RktSym{def}\mbox{\hphantom{\Scribtexttt{x}}}\RktVal{{\textquotesingle}}\RktVal{z}\mbox{\hphantom{\Scribtexttt{x}}}\RktVal{{\textquotesingle}}\RktVal{(}\RktVal{(}\RktVal{$\lambda$}\mbox{\hphantom{\Scribtexttt{x}}}\RktVal{x}\mbox{\hphantom{\Scribtexttt{x}}}\RktVal{={\Stttextmore}}\mbox{\hphantom{\Scribtexttt{x}}}\RktVal{(}\RktVal{$\lambda$}\mbox{\hphantom{\Scribtexttt{x}}}\RktVal{zf}\mbox{\hphantom{\Scribtexttt{x}}}\RktVal{={\Stttextmore}}\mbox{\hphantom{\Scribtexttt{x}}}\RktVal{(}\RktVal{$\lambda$}\mbox{\hphantom{\Scribtexttt{x}}}\RktVal{sf}\mbox{\hphantom{\Scribtexttt{x}}}\RktVal{={\Stttextmore}}\mbox{\hphantom{\Scribtexttt{x}}}\RktVal{zf}\RktVal{)}\RktVal{)}\RktVal{)}\mbox{\hphantom{\Scribtexttt{x}}}\RktVal{{\hbox{\texttt{:}}}}\mbox{\hphantom{\Scribtexttt{x}}}\RktVal{nat}\RktVal{)}\RktPn{)}

\RktPn{(}\RktSym{def}\mbox{\hphantom{\Scribtexttt{x}}}\RktVal{{\textquotesingle}}\RktVal{s}\mbox{\hphantom{\Scribtexttt{x}}}\RktVal{{\textquotesingle}}\RktVal{(}\RktVal{(}\RktVal{$\lambda$}\mbox{\hphantom{\Scribtexttt{x}}}\RktVal{n}\mbox{\hphantom{\Scribtexttt{x}}}\RktVal{={\Stttextmore}}\mbox{\hphantom{\Scribtexttt{x}}}\RktVal{(}\RktVal{$\lambda$}\mbox{\hphantom{\Scribtexttt{x}}}\RktVal{x}\mbox{\hphantom{\Scribtexttt{x}}}\RktVal{={\Stttextmore}}\mbox{\hphantom{\Scribtexttt{x}}}\RktVal{(}\RktVal{$\lambda$}\mbox{\hphantom{\Scribtexttt{x}}}\RktVal{zf}\mbox{\hphantom{\Scribtexttt{x}}}\RktVal{={\Stttextmore}}\mbox{\hphantom{\Scribtexttt{x}}}\RktVal{(}\RktVal{$\lambda$}\mbox{\hphantom{\Scribtexttt{x}}}\RktVal{sf}\mbox{\hphantom{\Scribtexttt{x}}}\RktVal{={\Stttextmore}}\mbox{\hphantom{\Scribtexttt{x}}}\RktVal{(}\RktVal{sf}\mbox{\hphantom{\Scribtexttt{x}}}\RktVal{(}\RktVal{(}\RktVal{(}\RktVal{n}\mbox{\hphantom{\Scribtexttt{x}}}\RktVal{x}\RktVal{)}\mbox{\hphantom{\Scribtexttt{x}}}\RktVal{zf}\RktVal{)}\mbox{\hphantom{\Scribtexttt{x}}}\RktVal{sf}\RktVal{)}\RktVal{)}\RktVal{)}\RktVal{)}\RktVal{)}\RktVal{)}\mbox{\hphantom{\Scribtexttt{x}}}\RktVal{{\hbox{\texttt{:}}}}\mbox{\hphantom{\Scribtexttt{x}}}\RktVal{(}\RktVal{nat}\mbox{\hphantom{\Scribtexttt{x}}}\RktVal{\mbox{{-}{\Stttextmore}}}\mbox{\hphantom{\Scribtexttt{x}}}\RktVal{nat}\RktVal{)}\RktVal{)}\RktPn{)}

\RktPn{(}\RktSym{def}\mbox{\hphantom{\Scribtexttt{x}}}\RktVal{{\textquotesingle}}\RktVal{one}\mbox{\hphantom{\Scribtexttt{x}}}\RktVal{{\textquotesingle}}\RktVal{(}\RktVal{(}\RktVal{s}\mbox{\hphantom{\Scribtexttt{x}}}\RktVal{z}\RktVal{)}\mbox{\hphantom{\Scribtexttt{x}}}\RktVal{{\hbox{\texttt{:}}}}\mbox{\hphantom{\Scribtexttt{x}}}\RktVal{nat}\RktVal{)}\RktPn{)}

\RktPn{(}\RktSym{def}\mbox{\hphantom{\Scribtexttt{x}}}\RktVal{{\textquotesingle}}\RktVal{two}\mbox{\hphantom{\Scribtexttt{x}}}\RktVal{{\textquotesingle}}\RktVal{(}\RktVal{(}\RktVal{s}\mbox{\hphantom{\Scribtexttt{x}}}\RktVal{(}\RktVal{s}\mbox{\hphantom{\Scribtexttt{x}}}\RktVal{z}\RktVal{)}\RktVal{)}\mbox{\hphantom{\Scribtexttt{x}}}\RktVal{{\hbox{\texttt{:}}}}\mbox{\hphantom{\Scribtexttt{x}}}\RktVal{nat}\RktVal{)}\RktPn{)}

\RktPn{(}\RktSym{def}\mbox{\hphantom{\Scribtexttt{x}}}\RktVal{{\textquotesingle}}\RktVal{plus}\mbox{\hphantom{\Scribtexttt{x}}}\RktVal{{\textquotesingle}}\RktVal{(}\RktVal{(}\RktVal{$\lambda$}\mbox{\hphantom{\Scribtexttt{x}}}\RktVal{x}\mbox{\hphantom{\Scribtexttt{x}}}\RktVal{={\Stttextmore}}\mbox{\hphantom{\Scribtexttt{x}}}\RktVal{(}\RktVal{$\lambda$}\mbox{\hphantom{\Scribtexttt{x}}}\RktVal{y}\mbox{\hphantom{\Scribtexttt{x}}}\RktVal{={\Stttextmore}}\mbox{\hphantom{\Scribtexttt{x}}}\RktVal{(}\RktVal{(}\RktVal{(}\RktVal{x}\mbox{\hphantom{\Scribtexttt{x}}}\RktVal{nat}\RktVal{)}\mbox{\hphantom{\Scribtexttt{x}}}\RktVal{y}\RktVal{)}\mbox{\hphantom{\Scribtexttt{x}}}\RktVal{s}\RktVal{)}\RktVal{)}\RktVal{)}\mbox{\hphantom{\Scribtexttt{x}}}\RktVal{{\hbox{\texttt{:}}}}\mbox{\hphantom{\Scribtexttt{x}}}\RktVal{(}\RktVal{nat}\mbox{\hphantom{\Scribtexttt{x}}}\RktVal{\mbox{{-}{\Stttextmore}}}\mbox{\hphantom{\Scribtexttt{x}}}\RktVal{(}\RktVal{nat}\mbox{\hphantom{\Scribtexttt{x}}}\RktVal{\mbox{{-}{\Stttextmore}}}\mbox{\hphantom{\Scribtexttt{x}}}\RktVal{nat}\RktVal{)}\RktVal{)}\RktVal{)}\RktPn{)}\end{SingleColumn}\end{RktBlk}

~

We can check that
\RktPn{((}\RktSym{plus~one}\RktPn{)}\RktSym{~one}\RktPn{)}
and \RktSym{two} are 
definitionally equivalent, but we have no way of expressing this in
our term language. The next step is to add equality. From this point on,
algorithmic descriptions are shorter and clearer than proof rules.

The proof term
\RktPn{(}\RktSym{eq-refl t}\RktPn{)} will have type
\RktPn{(}\RktSym{t = t}\RktPn{)}. The type
\RktPn{(}\RktSym{t = w}\RktPn{)} will have a proof term
exactly when \RktSym{t} and
\RktSym{w} are definitionally equal. The elimination form
\RktSym{eq-elim} implements the principle that ``equals may be
substituted for equals''. It is applied to five things: a term
\RktSym{t} of type \RktSym{T}, a ``property'' \RktSym{P} that has type
\RktPn{(}\RktSym{T -> Type}\RktPn{)},
a term \RktSym{pt} of type \RktPn{(}\RktSym{P t}\RktPn{)} (that
is, a proof that \RktSym{t} has property \RktSym{P}), a term
\RktSym{w} of type \RktSym{T}, and a term \RktSym{peq} of type
\RktPn{(}\RktSym{t = w}\RktPn{)}. The application of
\RktSym{eq-elim} has type \RktPn{(}\RktSym{P w}\RktPn{)}
(that is, it proves that
\RktSym{w} has property \RktSym{P}), and the reduction rule produces
the result of reducing \RktSym{pt}. All this takes more space to
describe in English than does the implementation. Here are some proofs.

~

\label{t:x28part_x28gentag_0x29x29}

\begin{RktBlk}\begin{SingleColumn}\RktPn{(}\RktSym{def}\mbox{\hphantom{\Scribtexttt{x}}}\RktVal{{\textquotesingle}}\RktVal{one{-}eq{-}one}\mbox{\hphantom{\Scribtexttt{x}}}\RktVal{{\textquotesingle}}\RktVal{(}\RktVal{(}\RktVal{eq{-}refl}\mbox{\hphantom{\Scribtexttt{x}}}\RktVal{one}\RktVal{)}\mbox{\hphantom{\Scribtexttt{x}}}\RktVal{{\hbox{\texttt{:}}}}\mbox{\hphantom{\Scribtexttt{x}}}\RktVal{(}\RktVal{one}\mbox{\hphantom{\Scribtexttt{x}}}\RktVal{=}\mbox{\hphantom{\Scribtexttt{x}}}\RktVal{one}\RktVal{)}\RktVal{)}\RktPn{)}

\RktPn{(}\RktSym{def}\mbox{\hphantom{\Scribtexttt{x}}}\RktVal{{\textquotesingle}}\RktVal{one{-}plus{-}one{-}is{-}two}\mbox{\hphantom{\Scribtexttt{x}}}\RktVal{{\textquotesingle}}\RktVal{(}\RktVal{(}\RktVal{eq{-}refl}\mbox{\hphantom{\Scribtexttt{x}}}\RktVal{two}\RktVal{)}\mbox{\hphantom{\Scribtexttt{x}}}\RktVal{{\hbox{\texttt{:}}}}\mbox{\hphantom{\Scribtexttt{x}}}\RktVal{(}\RktVal{(}\RktVal{(}\RktVal{plus}\mbox{\hphantom{\Scribtexttt{x}}}\RktVal{one}\RktVal{)}\mbox{\hphantom{\Scribtexttt{x}}}\RktVal{one}\RktVal{)}\mbox{\hphantom{\Scribtexttt{x}}}\RktVal{=}\mbox{\hphantom{\Scribtexttt{x}}}\RktVal{two}\RktVal{)}\RktVal{)}\RktPn{)}

\RktPn{(}\RktSym{def}\mbox{\hphantom{\Scribtexttt{x}}}\RktVal{{\textquotesingle}}\RktVal{eq{-}symm}

\mbox{\hphantom{\Scribtexttt{xx}}}\RktVal{{\textquotesingle}}\RktVal{(}\RktVal{(}\RktVal{$\lambda$}\mbox{\hphantom{\Scribtexttt{x}}}\RktVal{x}\mbox{\hphantom{\Scribtexttt{x}}}\RktVal{={\Stttextmore}}\mbox{\hphantom{\Scribtexttt{x}}}\RktVal{(}\RktVal{$\lambda$}\mbox{\hphantom{\Scribtexttt{x}}}\RktVal{y}\mbox{\hphantom{\Scribtexttt{x}}}\RktVal{={\Stttextmore}}\mbox{\hphantom{\Scribtexttt{x}}}\RktVal{(}\RktVal{$\lambda$}\mbox{\hphantom{\Scribtexttt{x}}}\RktVal{p}\mbox{\hphantom{\Scribtexttt{x}}}\RktVal{={\Stttextmore}}\mbox{\hphantom{\Scribtexttt{x}}}\RktVal{(}\RktVal{eq{-}elim}\mbox{\hphantom{\Scribtexttt{x}}}\RktVal{x}\mbox{\hphantom{\Scribtexttt{x}}}\RktVal{(}\RktVal{$\lambda$}\mbox{\hphantom{\Scribtexttt{x}}}\RktVal{w}\mbox{\hphantom{\Scribtexttt{x}}}\RktVal{={\Stttextmore}}\mbox{\hphantom{\Scribtexttt{x}}}\RktVal{(}\RktVal{w}\mbox{\hphantom{\Scribtexttt{x}}}\RktVal{=}\mbox{\hphantom{\Scribtexttt{x}}}\RktVal{x}\RktVal{)}\RktVal{)}\mbox{\hphantom{\Scribtexttt{x}}}\RktVal{(}\RktVal{eq{-}refl}\mbox{\hphantom{\Scribtexttt{x}}}\RktVal{x}\RktVal{)}\mbox{\hphantom{\Scribtexttt{x}}}\RktVal{y}\mbox{\hphantom{\Scribtexttt{x}}}\RktVal{p}\RktVal{)}\RktVal{)}\RktVal{)}\RktVal{)}

\mbox{\hphantom{\Scribtexttt{xxxxx}}}\RktVal{{\hbox{\texttt{:}}}}\mbox{\hphantom{\Scribtexttt{x}}}\RktVal{(}\RktVal{$\forall$}\mbox{\hphantom{\Scribtexttt{x}}}\RktVal{(}\RktVal{x}\mbox{\hphantom{\Scribtexttt{x}}}\RktVal{{\hbox{\texttt{:}}}}\mbox{\hphantom{\Scribtexttt{x}}}\RktVal{Type}\RktVal{)}\mbox{\hphantom{\Scribtexttt{x}}}\RktVal{\mbox{{-}{\Stttextmore}}}\mbox{\hphantom{\Scribtexttt{x}}}\RktVal{(}\RktVal{$\forall$}\mbox{\hphantom{\Scribtexttt{x}}}\RktVal{(}\RktVal{y}\mbox{\hphantom{\Scribtexttt{x}}}\RktVal{{\hbox{\texttt{:}}}}\mbox{\hphantom{\Scribtexttt{x}}}\RktVal{Type}\RktVal{)}\mbox{\hphantom{\Scribtexttt{x}}}\RktVal{\mbox{{-}{\Stttextmore}}}\mbox{\hphantom{\Scribtexttt{x}}}\RktVal{(}\RktVal{(}\RktVal{x}\mbox{\hphantom{\Scribtexttt{x}}}\RktVal{=}\mbox{\hphantom{\Scribtexttt{x}}}\RktVal{y}\RktVal{)}\mbox{\hphantom{\Scribtexttt{x}}}\RktVal{\mbox{{-}{\Stttextmore}}}\mbox{\hphantom{\Scribtexttt{x}}}\RktVal{(}\RktVal{y}\mbox{\hphantom{\Scribtexttt{x}}}\RktVal{=}\mbox{\hphantom{\Scribtexttt{x}}}\RktVal{x}\RktVal{)}\RktVal{)}\RktVal{)}\RktVal{)}\RktVal{)}\RktPn{)}

\RktPn{(}\RktSym{def}\mbox{\hphantom{\Scribtexttt{x}}}\RktVal{{\textquotesingle}}\RktVal{eq{-}trans}

\mbox{\hphantom{\Scribtexttt{xx}}}\RktVal{{\textquotesingle}}\RktVal{(}\RktVal{(}\RktVal{$\lambda$}\mbox{\hphantom{\Scribtexttt{x}}}\RktVal{x}\mbox{\hphantom{\Scribtexttt{x}}}\RktVal{={\Stttextmore}}\mbox{\hphantom{\Scribtexttt{x}}}\RktVal{(}\RktVal{$\lambda$}\mbox{\hphantom{\Scribtexttt{x}}}\RktVal{y}\mbox{\hphantom{\Scribtexttt{x}}}\RktVal{={\Stttextmore}}\mbox{\hphantom{\Scribtexttt{x}}}\RktVal{(}\RktVal{$\lambda$}\mbox{\hphantom{\Scribtexttt{x}}}\RktVal{z}\mbox{\hphantom{\Scribtexttt{x}}}\RktVal{={\Stttextmore}}\mbox{\hphantom{\Scribtexttt{x}}}\RktVal{(}\RktVal{$\lambda$}\mbox{\hphantom{\Scribtexttt{x}}}\RktVal{p}\mbox{\hphantom{\Scribtexttt{x}}}\RktVal{={\Stttextmore}}\mbox{\hphantom{\Scribtexttt{x}}}\RktVal{(}\RktVal{$\lambda$}\mbox{\hphantom{\Scribtexttt{x}}}\RktVal{q}\mbox{\hphantom{\Scribtexttt{x}}}\RktVal{={\Stttextmore}}\mbox{\hphantom{\Scribtexttt{x}}}\RktVal{(}\RktVal{eq{-}elim}\mbox{\hphantom{\Scribtexttt{x}}}\RktVal{y}\mbox{\hphantom{\Scribtexttt{x}}}\RktVal{(}\RktVal{$\lambda$}\mbox{\hphantom{\Scribtexttt{x}}}\RktVal{w}\mbox{\hphantom{\Scribtexttt{x}}}\RktVal{={\Stttextmore}}\mbox{\hphantom{\Scribtexttt{x}}}\RktVal{(}\RktVal{x}\mbox{\hphantom{\Scribtexttt{x}}}\RktVal{=}\mbox{\hphantom{\Scribtexttt{x}}}\RktVal{w}\RktVal{)}\RktVal{)}\mbox{\hphantom{\Scribtexttt{x}}}\RktVal{p}\mbox{\hphantom{\Scribtexttt{x}}}\RktVal{z}\mbox{\hphantom{\Scribtexttt{x}}}\RktVal{q}\RktVal{)}\RktVal{)}\RktVal{)}\RktVal{)}\RktVal{)}\RktVal{)}

\mbox{\hphantom{\Scribtexttt{xxxxx}}}\RktVal{{\hbox{\texttt{:}}}}\mbox{\hphantom{\Scribtexttt{x}}}\RktVal{(}\RktVal{$\forall$}\mbox{\hphantom{\Scribtexttt{x}}}\RktVal{(}\RktVal{x}\mbox{\hphantom{\Scribtexttt{x}}}\RktVal{{\hbox{\texttt{:}}}}\mbox{\hphantom{\Scribtexttt{x}}}\RktVal{Type}\RktVal{)}\mbox{\hphantom{\Scribtexttt{x}}}\RktVal{\mbox{{-}{\Stttextmore}}}\mbox{\hphantom{\Scribtexttt{x}}}\RktVal{(}\RktVal{$\forall$}\mbox{\hphantom{\Scribtexttt{x}}}\RktVal{(}\RktVal{y}\mbox{\hphantom{\Scribtexttt{x}}}\RktVal{{\hbox{\texttt{:}}}}\mbox{\hphantom{\Scribtexttt{x}}}\RktVal{Type}\RktVal{)}\mbox{\hphantom{\Scribtexttt{x}}}\RktVal{\mbox{{-}{\Stttextmore}}}\mbox{\hphantom{\Scribtexttt{x}}}\RktVal{(}\RktVal{$\forall$}\mbox{\hphantom{\Scribtexttt{x}}}\RktVal{(}\RktVal{z}\mbox{\hphantom{\Scribtexttt{x}}}\RktVal{{\hbox{\texttt{:}}}}\mbox{\hphantom{\Scribtexttt{x}}}\RktVal{Type}\RktVal{)}\mbox{\hphantom{\Scribtexttt{x}}}\RktVal{\mbox{{-}{\Stttextmore}}}

\mbox{\hphantom{\Scribtexttt{xxxxxxxxxxxxxxxxxxxxxxxxxxxxx}}}\RktVal{(}\RktVal{(}\RktVal{x}\mbox{\hphantom{\Scribtexttt{x}}}\RktVal{=}\mbox{\hphantom{\Scribtexttt{x}}}\RktVal{y}\RktVal{)}\mbox{\hphantom{\Scribtexttt{x}}}\RktVal{\mbox{{-}{\Stttextmore}}}\mbox{\hphantom{\Scribtexttt{x}}}\RktVal{(}\RktVal{(}\RktVal{y}\mbox{\hphantom{\Scribtexttt{x}}}\RktVal{=}\mbox{\hphantom{\Scribtexttt{x}}}\RktVal{z}\RktVal{)}\mbox{\hphantom{\Scribtexttt{x}}}\RktVal{\mbox{{-}{\Stttextmore}}}\mbox{\hphantom{\Scribtexttt{x}}}\RktVal{(}\RktVal{x}\mbox{\hphantom{\Scribtexttt{x}}}\RktVal{=}\mbox{\hphantom{\Scribtexttt{x}}}\RktVal{z}\RktVal{)}\RktVal{)}\RktVal{)}\RktVal{)}\RktVal{)}\RktVal{)}\RktVal{)}\RktPn{)}\end{SingleColumn}\end{RktBlk}

~

The Church numerals are inefficient and awkward. Furthermore, even if
we add back $\bot$ (which we should do, to facilitate expressing logical
negation), we can state but cannot prove $\lnot (0=1)$. To address
these issues, we implement Peano numbers, with type \RktSym{Nat}, zero
\RktSym{Z} and successor function \RktSym{S}.  What should the
elimination form be? Computationally, we use natural numbers via
structural recursion, with a \RktSym{Z} case and a function to be
applied to the predecessor in the \RktSym{S} case. We could implement
this with a recursor \RktSym{nat-rec}, whose type would be $\forall(T:
Type) \ra T \ra (T \ra T) \ra \mathit{Nat} \ra T$.  The reduction
rules would reduce \RktPn{(}\RktSym{nat-rec ZCase SCase 0}\RktPn{)} to
\RktSym{ZCase} and would reduce
\RktPn{(}\RktSym{nat-rec ZCase SCase~}\RktPn{(}\RktSym{S n}\RktPn{))}
to \RktPn{(}\RktSym{SCase~}\RktPn{(}\RktSym{nat-rec ZCase SCase n}\RktPn{))}.

But we can generalize $T$ to be dependent, having it be a property $P$
indexed by a $\mathit{Nat}$. The type becomes
$\forall (P : Nat\ra Type) \ra (P~Z) \ra 
(\forall (k:Nat) \ra (P~k \ra P~(S~k))) \ra
(\forall (k: Nat) \ra P~k)$. This is familiar: it is the induction
principle for natural numbers, which we can call \RktSym{nat-ind}.
If $P$ is the constant function that
produces $T$, we recover the recursor, and the reduction rules for
\RktSym{nat-ind} are the obvious generalization of those for \RktSym{nat-rec}.

\clearpage

\label{t:x28part_x28gentag_0x29x29}

\begin{RktBlk}\begin{SingleColumn}\RktPn{(}\RktSym{def}\mbox{\hphantom{\Scribtexttt{x}}}\RktVal{{\textquotesingle}}\RktVal{nat{-}rec}

\mbox{\hphantom{\Scribtexttt{xx}}}\RktVal{{\textquotesingle}}\RktVal{(}\RktVal{(}\RktVal{$\lambda$}\mbox{\hphantom{\Scribtexttt{x}}}\RktVal{C}\mbox{\hphantom{\Scribtexttt{x}}}\RktVal{={\Stttextmore}}\mbox{\hphantom{\Scribtexttt{x}}}\RktVal{(}\RktVal{$\lambda$}\mbox{\hphantom{\Scribtexttt{x}}}\RktVal{zc}\mbox{\hphantom{\Scribtexttt{x}}}\RktVal{={\Stttextmore}}\mbox{\hphantom{\Scribtexttt{x}}}\RktVal{(}\RktVal{$\lambda$}\mbox{\hphantom{\Scribtexttt{x}}}\RktVal{sc}\mbox{\hphantom{\Scribtexttt{x}}}\RktVal{={\Stttextmore}}\mbox{\hphantom{\Scribtexttt{x}}}\RktVal{(}\RktVal{$\lambda$}\mbox{\hphantom{\Scribtexttt{x}}}\RktVal{n}\mbox{\hphantom{\Scribtexttt{x}}}\RktVal{={\Stttextmore}}\mbox{\hphantom{\Scribtexttt{x}}}\RktVal{(}\RktVal{nat{-}ind}\mbox{\hphantom{\Scribtexttt{x}}}\RktVal{(}\RktVal{$\lambda$}\mbox{\hphantom{\Scribtexttt{x}}}\RktVal{{\char`\_}}\mbox{\hphantom{\Scribtexttt{x}}}\RktVal{={\Stttextmore}}\mbox{\hphantom{\Scribtexttt{x}}}\RktVal{C}\RktVal{)}\mbox{\hphantom{\Scribtexttt{x}}}\RktVal{zc}\mbox{\hphantom{\Scribtexttt{x}}}\RktVal{(}\RktVal{$\lambda$}\mbox{\hphantom{\Scribtexttt{x}}}\RktVal{{\char`\_}}\mbox{\hphantom{\Scribtexttt{x}}}\RktVal{={\Stttextmore}}\mbox{\hphantom{\Scribtexttt{x}}}\RktVal{sc}\RktVal{)}\mbox{\hphantom{\Scribtexttt{x}}}\RktVal{n}\RktVal{)}\RktVal{)}\RktVal{)}\RktVal{)}\RktVal{)}

\mbox{\hphantom{\Scribtexttt{xxxx}}}\RktVal{{\hbox{\texttt{:}}}}\mbox{\hphantom{\Scribtexttt{x}}}\RktVal{(}\RktVal{$\forall$}\mbox{\hphantom{\Scribtexttt{x}}}\RktVal{(}\RktVal{C}\mbox{\hphantom{\Scribtexttt{x}}}\RktVal{{\hbox{\texttt{:}}}}\mbox{\hphantom{\Scribtexttt{x}}}\RktVal{Type}\RktVal{)}\mbox{\hphantom{\Scribtexttt{x}}}\RktVal{\mbox{{-}{\Stttextmore}}}\mbox{\hphantom{\Scribtexttt{x}}}\RktVal{(}\RktVal{C}\mbox{\hphantom{\Scribtexttt{x}}}\RktVal{\mbox{{-}{\Stttextmore}}}\mbox{\hphantom{\Scribtexttt{x}}}\RktVal{(}\RktVal{C}\mbox{\hphantom{\Scribtexttt{x}}}\RktVal{\mbox{{-}{\Stttextmore}}}\mbox{\hphantom{\Scribtexttt{x}}}\RktVal{C}\RktVal{)}\mbox{\hphantom{\Scribtexttt{x}}}\RktVal{\mbox{{-}{\Stttextmore}}}\mbox{\hphantom{\Scribtexttt{x}}}\RktVal{Nat}\mbox{\hphantom{\Scribtexttt{x}}}\RktVal{\mbox{{-}{\Stttextmore}}}\mbox{\hphantom{\Scribtexttt{x}}}\RktVal{C}\RktVal{)}\RktVal{)}\RktVal{)}\RktPn{)}

\mbox{\hphantom{\Scribtexttt{x}}}

\RktPn{(}\RktSym{def}\mbox{\hphantom{\Scribtexttt{x}}}\RktVal{{\textquotesingle}}\RktVal{plus}

\mbox{\hphantom{\Scribtexttt{xx}}}\RktVal{{\textquotesingle}}\RktVal{(}\RktVal{(}\RktVal{$\lambda$}\mbox{\hphantom{\Scribtexttt{x}}}\RktVal{n}\mbox{\hphantom{\Scribtexttt{x}}}\RktVal{={\Stttextmore}}\mbox{\hphantom{\Scribtexttt{x}}}\RktVal{(}\RktVal{nat{-}rec}\mbox{\hphantom{\Scribtexttt{x}}}\RktVal{(}\RktVal{Nat}\mbox{\hphantom{\Scribtexttt{x}}}\RktVal{\mbox{{-}{\Stttextmore}}}\mbox{\hphantom{\Scribtexttt{x}}}\RktVal{Nat}\RktVal{)}\mbox{\hphantom{\Scribtexttt{x}}}\RktVal{(}\RktVal{$\lambda$}\mbox{\hphantom{\Scribtexttt{x}}}\RktVal{m}\mbox{\hphantom{\Scribtexttt{x}}}\RktVal{={\Stttextmore}}\mbox{\hphantom{\Scribtexttt{x}}}\RktVal{m}\RktVal{)}\mbox{\hphantom{\Scribtexttt{x}}}\RktVal{(}\RktVal{$\lambda$}\mbox{\hphantom{\Scribtexttt{x}}}\RktVal{pm}\mbox{\hphantom{\Scribtexttt{x}}}\RktVal{={\Stttextmore}}\mbox{\hphantom{\Scribtexttt{x}}}\RktVal{(}\RktVal{$\lambda$}\mbox{\hphantom{\Scribtexttt{x}}}\RktVal{x}\mbox{\hphantom{\Scribtexttt{x}}}\RktVal{={\Stttextmore}}\mbox{\hphantom{\Scribtexttt{x}}}\RktVal{(}\RktVal{S}\mbox{\hphantom{\Scribtexttt{x}}}\RktVal{(}\RktVal{pm}\mbox{\hphantom{\Scribtexttt{x}}}\RktVal{x}\RktVal{)}\RktVal{)}\RktVal{)}\RktVal{)}\mbox{\hphantom{\Scribtexttt{x}}}\RktVal{n}\RktVal{)}\RktVal{)}

\mbox{\hphantom{\Scribtexttt{xxxx}}}\RktVal{{\hbox{\texttt{:}}}}\mbox{\hphantom{\Scribtexttt{x}}}\RktVal{(}\RktVal{Nat}\mbox{\hphantom{\Scribtexttt{x}}}\RktVal{\mbox{{-}{\Stttextmore}}}\mbox{\hphantom{\Scribtexttt{x}}}\RktVal{Nat}\mbox{\hphantom{\Scribtexttt{x}}}\RktVal{\mbox{{-}{\Stttextmore}}}\mbox{\hphantom{\Scribtexttt{x}}}\RktVal{Nat}\RktVal{)}\RktVal{)}\RktPn{)}

\mbox{\hphantom{\Scribtexttt{x}}}

\RktPn{(}\RktSym{def}\mbox{\hphantom{\Scribtexttt{x}}}\RktVal{{\textquotesingle}}\RktVal{plus{-}zero{-}left}\mbox{\hphantom{\Scribtexttt{x}}}\RktVal{{\textquotesingle}}\RktVal{(}\RktVal{(}\RktVal{$\lambda$}\mbox{\hphantom{\Scribtexttt{x}}}\RktVal{n}\mbox{\hphantom{\Scribtexttt{x}}}\RktVal{={\Stttextmore}}\mbox{\hphantom{\Scribtexttt{x}}}\RktVal{(}\RktVal{eq{-}refl}\mbox{\hphantom{\Scribtexttt{x}}}\RktVal{n}\RktVal{)}\RktVal{)}\mbox{\hphantom{\Scribtexttt{x}}}\RktVal{{\hbox{\texttt{:}}}}\mbox{\hphantom{\Scribtexttt{x}}}\RktVal{(}\RktVal{$\forall$}\mbox{\hphantom{\Scribtexttt{x}}}\RktVal{(}\RktVal{n}\mbox{\hphantom{\Scribtexttt{x}}}\RktVal{{\hbox{\texttt{:}}}}\mbox{\hphantom{\Scribtexttt{x}}}\RktVal{Nat}\RktVal{)}\mbox{\hphantom{\Scribtexttt{x}}}\RktVal{\mbox{{-}{\Stttextmore}}}\mbox{\hphantom{\Scribtexttt{x}}}\RktVal{(}\RktVal{(}\RktVal{plus}\mbox{\hphantom{\Scribtexttt{x}}}\RktVal{Z}\mbox{\hphantom{\Scribtexttt{x}}}\RktVal{n}\RktVal{)}\mbox{\hphantom{\Scribtexttt{x}}}\RktVal{=}\mbox{\hphantom{\Scribtexttt{x}}}\RktVal{n}\RktVal{)}\RktVal{)}\RktVal{)}\RktPn{)}\end{SingleColumn}\end{RktBlk}

\label{t:x28part_x28gentag_0x29x29}

\begin{RktBlk}\begin{SingleColumn}\RktPn{(}\RktSym{def}\mbox{\hphantom{\Scribtexttt{x}}}\RktVal{{\textquotesingle}}\RktVal{plus{-}zero{-}right}

\mbox{\hphantom{\Scribtexttt{xx}}}\RktVal{{\textquotesingle}}\RktVal{(}\RktVal{(}\RktVal{$\lambda$}\mbox{\hphantom{\Scribtexttt{x}}}\RktVal{n}\mbox{\hphantom{\Scribtexttt{x}}}\RktVal{={\Stttextmore}}\mbox{\hphantom{\Scribtexttt{x}}}\RktVal{(}\RktVal{nat{-}ind}\mbox{\hphantom{\Scribtexttt{x}}}\RktVal{(}\RktVal{$\lambda$}\mbox{\hphantom{\Scribtexttt{x}}}\RktVal{m}\mbox{\hphantom{\Scribtexttt{x}}}\RktVal{={\Stttextmore}}\mbox{\hphantom{\Scribtexttt{x}}}\RktVal{(}\RktVal{(}\RktVal{plus}\mbox{\hphantom{\Scribtexttt{x}}}\RktVal{m}\mbox{\hphantom{\Scribtexttt{x}}}\RktVal{Z}\RktVal{)}\mbox{\hphantom{\Scribtexttt{x}}}\RktVal{=}\mbox{\hphantom{\Scribtexttt{x}}}\RktVal{m}\RktVal{)}\RktVal{)}\mbox{\hphantom{\Scribtexttt{x}}}\RktVal{(}\RktVal{eq{-}refl}\mbox{\hphantom{\Scribtexttt{x}}}\RktVal{Z}\RktVal{)}

\mbox{\hphantom{\Scribtexttt{xxxxxxxxxxxxxxxxxxxxx}}}\RktVal{(}\RktVal{$\lambda$}\mbox{\hphantom{\Scribtexttt{x}}}\RktVal{k}\mbox{\hphantom{\Scribtexttt{x}}}\RktVal{={\Stttextmore}}\mbox{\hphantom{\Scribtexttt{x}}}\RktVal{(}\RktVal{$\lambda$}\mbox{\hphantom{\Scribtexttt{x}}}\RktVal{p}\mbox{\hphantom{\Scribtexttt{x}}}\RktVal{={\Stttextmore}}

\mbox{\hphantom{\Scribtexttt{xxxxxxxxxxxxxxxxxxxxxxx}}}\RktVal{(}\RktVal{eq{-}elim}\mbox{\hphantom{\Scribtexttt{x}}}\RktVal{(}\RktVal{plus}\mbox{\hphantom{\Scribtexttt{x}}}\RktVal{k}\mbox{\hphantom{\Scribtexttt{x}}}\RktVal{Z}\RktVal{)}\mbox{\hphantom{\Scribtexttt{x}}}\RktVal{(}\RktVal{$\lambda$}\mbox{\hphantom{\Scribtexttt{x}}}\RktVal{w}\mbox{\hphantom{\Scribtexttt{x}}}\RktVal{={\Stttextmore}}\mbox{\hphantom{\Scribtexttt{x}}}\RktVal{(}\RktVal{(}\RktVal{S}\mbox{\hphantom{\Scribtexttt{x}}}\RktVal{(}\RktVal{plus}\mbox{\hphantom{\Scribtexttt{x}}}\RktVal{k}\mbox{\hphantom{\Scribtexttt{x}}}\RktVal{Z}\RktVal{)}\RktVal{)}\mbox{\hphantom{\Scribtexttt{x}}}\RktVal{=}\mbox{\hphantom{\Scribtexttt{x}}}\RktVal{(}\RktVal{S}\mbox{\hphantom{\Scribtexttt{x}}}\RktVal{w}\RktVal{)}\RktVal{)}\RktVal{)}

\mbox{\hphantom{\Scribtexttt{xxxxxxxxxxxxxxxxxxxxxxxxxxxxxxxx}}}\RktVal{(}\RktVal{eq{-}refl}\mbox{\hphantom{\Scribtexttt{x}}}\RktVal{(}\RktVal{S}\mbox{\hphantom{\Scribtexttt{x}}}\RktVal{(}\RktVal{plus}\mbox{\hphantom{\Scribtexttt{x}}}\RktVal{k}\mbox{\hphantom{\Scribtexttt{x}}}\RktVal{Z}\RktVal{)}\RktVal{)}\RktVal{)}\mbox{\hphantom{\Scribtexttt{x}}}\RktVal{k}\mbox{\hphantom{\Scribtexttt{x}}}\RktVal{p}\RktVal{)}\RktVal{)}\RktVal{)}

\mbox{\hphantom{\Scribtexttt{xxxxxxxxxxxxxxxxxxxxx}}}\RktVal{n}\RktVal{)}\RktVal{)}

\mbox{\hphantom{\Scribtexttt{xxxxx}}}\RktVal{{\hbox{\texttt{:}}}}\mbox{\hphantom{\Scribtexttt{x}}}\RktVal{(}\RktVal{$\forall$}\mbox{\hphantom{\Scribtexttt{x}}}\RktVal{(}\RktVal{n}\mbox{\hphantom{\Scribtexttt{x}}}\RktVal{{\hbox{\texttt{:}}}}\mbox{\hphantom{\Scribtexttt{x}}}\RktVal{Nat}\RktVal{)}\mbox{\hphantom{\Scribtexttt{x}}}\RktVal{\mbox{{-}{\Stttextmore}}}\mbox{\hphantom{\Scribtexttt{x}}}\RktVal{(}\RktVal{(}\RktVal{plus}\mbox{\hphantom{\Scribtexttt{x}}}\RktVal{n}\mbox{\hphantom{\Scribtexttt{x}}}\RktVal{Z}\RktVal{)}\mbox{\hphantom{\Scribtexttt{x}}}\RktVal{=}\mbox{\hphantom{\Scribtexttt{x}}}\RktVal{n}\RktVal{)}\RktVal{)}\RktVal{)}\RktPn{)}\end{SingleColumn}\end{RktBlk}

~

That last definition demonstrates the capability we promised at the
start of this section, and we can also prove $\lnot(0=1)$, or more
generally, ${\forall(n:\mathit{Nat})\ra \lnot(0=S~n)}$. To fulfil the
promise of the title of the section, we need to add the existential
quantifier. We use the notation $\exists(x:T)\ra W$ for the dependent
sum type often called $\Sigma$ in the literature (again, see the title of
\cite{ADLO}). The introduction form \RktSym{$\exists$-intro} tuples
a type $T$, a ``witness'' $a$ of type $T$, and a value
$\mathit{pa}$ of type $W[x\mapsto a]$ (that is, a proof that $a$ has
property $W$). The elimination form \RktSym{$\exists$-elim} is
applied to a type $V$, a function $f$ of type $\forall(x:T)\ra(W\ra
V)$, and a value $b$ of type $\exists(x:T)\ra W$. The witness $a$
embedded in $b$ will be used as the argument to $f$ to produce a value
of type $V$, which is the type of the application of the elimination
form. The associated reduction rule reduces
\RktPn{(}\RktSym{$\exists$-elim V f~}\RktPn{(}\RktSym{$\exists$-intro
  T a pa}\RktPn{)}\RktPn{)} to \RktPn{(}\RktSym{f a pa}\RktPn{)}.

As was the situation in the previous section, we have implemented
intuitionistic logic, and so there are some theorems of classical
logic that we cannot prove. For example, we can prove the theorem
$\lnot (\exists (x:A)\ra B) \ra (\forall (x:A)\ra \lnot B)$, but 
$\lnot (\forall (x:A)\ra B) \ra (\exists (x:A)\ra \lnot B)$ eludes us.
By adding the law of the excluded middle, suitably quantified, as an
antecedent, we can prove the latter.

Why stop here? The code base is still under three hundred lines, and there are
some easy additions that make Proust more friendly. The encodings of
$\land$ and $\lor$ are not inefficient, but they are awkward to use,
and it is simple to add them as we did in section \ref{Prop}. We could
add lists as a datatype without much difficulty. Racket supports
various UI improvements. But many technical improvements significantly
complicate matters: a proper treatment of holes in the context of
dependent types; dependent pattern matching to simplify the use of
eliminators; a general mechanism for user-defined recursive datatypes;
a tactics language to avoid large explicit proof terms.

This is a natural transition point from Proust. Students are ready to
move to a system like Agda or Coq. Agda uses explicit proof terms
(eased by pattern matching and a nice Emacs interface); Coq hides
proof terms, though they can be exposed and even written explicitly as
needed. Proust's treatment of holes, quantifiers, typechecking, and
induction principles are in line with these more sophisticated
systems. We have fulfilled our twin goals of demystification of
full-featured proof assistants and properly motivated/situated
exposure to logic for computer science students. 

\section{Curriculum}\label{Curric}

The previous two sections focussed on the development of Proust as it
might be presented to students, but a course using it will intersperse
attention to related topics. A nonstandard treatment of a subject must
take care to demonstrate connections to more traditional approaches,
and that will be a major theme in this section.

The main point of the syntactic notion of proof is to be able to draw
conclusions about the semantic notion of truth. It is standard to write
$\Gamma\vdash T$ if there is a proof of $T$ using $\Gamma$ (in our
notation, if there exists $t$ such that $\Gamma\vdash t:T$).
Students will be familiar with Boolean values and functions from
programming, and have a tendency to confuse proof and truth, so it is
best to introduce early the formal definition of a valuation for
variables, the value of a formula with respect to a valuation, and the
notation $\Gamma\vDash T$ if every valuation that makes the formulas
in $\Gamma$ true also makes $T$ true.

One can then discuss the notion of soundness ($\Gamma\vdash T$ always
implies $\Gamma\vDash T$) and completeness (the converse). For a CS
student, soundness is the more useful property; completeness is
interesting more from a metamathematical viewpoint.
Our logics are sound, but they are not complete.
It is worth mentioning the existence of alternate
models for which our logics are complete (Heyting algebras, Kripke models),
and perhaps demonstrating the use of these for unprovability results.

My experience of requiring students to write natural deduction proofs
on paper is that they are frustrated
by the inflexibility of the formal system relative to their previous
experiences with informal proofs and what they know of Boolean
algebra. While a buggy program may also frustrate them, they are more
inclined to see it as something worth fixing. With Proust, an
incorrect proof is a buggy program, and this combined with immediate
feedback on correctness puts proof construction into familiar
territory. Students should still see a couple of proofs presented as
trees built from rules in the standard fashion, and the explicit
correspondence with proof terms.

In the bidirectional system we use, an immediate application of a
lambda cannot be typechecked; proofs must be normal forms. But
students may know from earlier courses
that immediate application of lambda is the
easiest way to get local definitions (as in the Racket macro
implementation of the \RktSym{let} construct). Proofs in propositional
logic are usually not complicated enough to require local definitions,
and if necessary, either the definition mechanism of section
\ref{Pred} can be implemented earlier, or a specific local binding
construct can be added to the term language.

The treatment of propositional logic in section \ref{Prop},
is close to a traditional approach. The main differences
are the restriction to intuitionistic logic and the use of Proust.
Lectures should discuss the lack of intuition and simple computational
interpretations for rules that result in classical logic
(such as double negation elimination), 
and that they can be added as axioms to Proust without difficulty.

The treatment of predicate logic in section \ref{Pred} is not at all
traditional, and students will benefit from a sketch of first-order
logic, notably the rules and axioms needed to add equality and
arithmetic, and the corresponding completeness and incompleteness
results.

I included the Church encodings above to demonstrate the expressivity
of the type system, but this material may not be suitable for all students.
As an alternative, one can easily add back to Proust the mechanisms developed
in the section of propositional logic.

It is not clear how much time there will be, in a one-semester
treatment, for students to spend much time with Agda or Coq. At the
least, a demonstration should be included at the end, and perhaps
a short one at the beginning to increase motivation. 

Finally, students will benefit from an understanding of the historical
development of this material. It is best incorporated in short
vignettes through the term, rather than crammed into a history lecture
full of names and dates.

At the time of publication, the first offering of this version of L\&C
is in progress as a special enriched section in the Fall 2016 term.
I hope to use this material with regular sections in Spring 2017.
The material could be offered at the introductory graduate level,
at an increased pace that
permits the inclusion of additional material (System F, G\"odel's
System T) or more exposure to Agda and/or Coq. The section on
predicate logic could be adapted as an introductory or parallel
tutorial for a graduate course that makes heavy use of these proof
assistants. The advantage of Proust over the tutorials cited earlier
is primarily due to the low overhead of Racket and an S-expression
representation of proofs and logical statements, the ease of minor uses
of mutation (counters, hash tables), and the advantages of
the DrRacket IDE. It is possible that use of a dynamically-typed
language to implement typechecking may avoid student confusion between
levels of abstraction.

\section{Acknowledgments}

I would like to thank Stephanie Weirich for her prompt and helpful
answers to my questions, the Recurse Center in New York City for
hosting me as a resident in October 2015 (during which time I
workshopped some of this material), and the anonymous referees
for their helpful suggestions.

\nocite{*}
\bibliographystyle{eptcs}
\bibliography{tfpie}
\end{document}